\documentclass[12pt]{article}
\usepackage{amsfonts}
\usepackage{amssymb, amsthm, amsfonts}
\usepackage{graphicx}
\usepackage[leqno]{amsmath}
\usepackage[normal]{subfigure}

\theoremstyle{plain}
\newtheorem{theorem}{{\Large T}HEOREM}
\newtheorem{proposition}{{\Large P}ROPORSITION}

\theoremstyle{definition}

\renewcommand{\theequation}{\arabic{section}.\arabic{equation}}
\renewcommand{\theequation}{\thesection\arabic{equation}}
\makeatletter
\renewcommand\thesection {\@arabic\c@section .}

\makeatletter
\usepackage{indentfirst}

\begin{document}

\title{{\bf Insider Trading in the Market with Rational Expected Price }}
\author{{\Large B}Y \quad {\Large F}UZHOU {\Large G}ONG, \quad
{\Large D}EQING {\Large Z}HOU \footnote{For useful discussions, we
thank Hong Liu, Yonghong Liu and Quanli Qin. We are grateful for
financial support for National Natural Science Foundation of China
(No.10721101), China's National 973 Project (No.2006CB805900) and
985 Project of Business Statistics and Econometrics Platform of
Peking University (No.468-11807-001).}}
\date{}
\makeatletter
\let\@fnsymbola=\@fnsymbol
\let\@fnsymbol=\@arabic
\maketitle
\let\@fnsymbol=\@fnsymbola
\let\@fnsymbola\relax
\makeatother
 \par \vskip .3cm \noindent

Kyle (1985) builds a pioneering and influential model, in which an
insider with long-lived private information submits an optimal order
in each period given the market maker's pricing rule. An
inconsistency exists to some extent in the sense that the ``constant
pricing rule " actually assumes an adaptive expected price with
pricing rule given before insider making the decision, and the
``market efficiency" condition, however, assumes a rational expected
price and implies that the pricing rule can be influenced by
insider's strategy. We loosen the ``constant pricing rule "
assumption by taking into account sufficiently the insider's
strategy has on pricing rule. According to the characteristic of the
conditional expectation of the informed profits, three different
models vary with insider's attitudes regarding to risk are
presented. Compared to Kyle (1985), the risk-averse insider in Model
1 can obtain larger guaranteed profits, the risk-neutral insider in
Model 2 can obtain a larger ex ante expectation of total profits
across all periods and the risk-seeking insider in Model 3  can
obtain larger risky profits. Moreover, the limit behaviors of the
three models when trading frequency approaches infinity are given,
showing that Model 1 acquires a strong-form efficiency, Model 2
acquires the Kyle's (1985) continuous equilibrium, and Model 3
acquires an equilibrium with information released at an increasing
speed.

 \vspace{0.3cm}

 {K{\scriptsize EYWORDS}:  Kyle (1985) model;
private information; pricing rule. \vspace{0.3cm}

 \begin{center}\section{\small{ INTRODUCTION}}\end{center}

  \noindent {\large T}HE MOTIVATION of our paper is to improve the canonical strategic
 trading model due to Kyle (1985). In  Kyle (1985), the price adjustment made by market maker
 is proportional to the total trading volume in which the proportional coefficient $\lambda_{n}$,
named as the pricing rule (or liquidity parameter, or inverse of
market depth), reflects the market maker's sensitivity regarding to
the  total trading volume. A strong assumption is the ``constant
pricing rule", which means that the insider takes the pricing rule
as a constant and thus ignores the effect her strategy has on it. We
loosen this assumption by taking into account sufficiently the
 effect that insider's choice might have on pricing rule.

 The ``constant pricing rule"  is not only just a ``very
 strong" assumption, but also can induce to some extent an inconsistency in
 Kyle (1985). In fact,
 the (semi-strong) market efficiency condition
 implies that the  pricing rule  $\lambda_{n}$
 dose can be affected by insider's submission. In other words,
the constant pricing rule announced by market maker is untrustable
in an semi-strong efficient market. Accordingly, the insider has an
incentive to deviate from the optimal strategy depicted in Kyle
(1985) to make a more profitable strategy since she know the market
maker would adjust the price to the deviated strategy to satisfy the
market efficiency condition. Thus, a new equilibrium arises in which
the insider's strategy can be characterized in a more
 reasonable manner.

An interesting specific case is the one period Kyle model, where
equilibrium is the same whether with the constant pricing rule
assumption or not. Admati and Pfleiderer (1988) point out this
coincidence when they investigate the clustering phenomena in a
model with short lived private information. Generally, in each
period except the last one in a multiple periods model, as long as
the insider takes market maker's response into account, the
conditional expectation of profits over the remaining periods, as a
random variance, has no maximum any more since the risky profits and
guaranteed profits that constitute the conditional expectation
cannot attain their maximums simultaneously. Hence we present three
different models varies with the maximization manner. Model 1
focuses on the risk-averse insider who maximizes the guaranteed
profits firstly and then, if multiple solutions are obtained,
chooses among them the one that maximizes the risky profits. Insider
in Model 2 is risk-neutral, trying to maximize the ex ante
expectation of total profits. While Model 3 assumes a risk-seeking
insider who maximizes the profits in an order reverse to that in
Model 1.

In Model 1 with a risk-averse insider, when trading happens
indefinitely frequently, the private information is incorporated
into price almost immediately, thus in limit Model 1 presents a
``strong efficient" market that defined by Fama (1970) as one with
prices reflecting both public and private information. This result
is analogous to Holden and Subrahamanyam (1992) with multiple
perfectly informed traders. However, there is only one insider in
Model 1 and the source of this result is risk aversion, not like
theirs - the aggressive competition among insiders. Chau and Vayanos
(2008) also obtain a strong form efficiency with one insider when
trading happens frequently. Their conclusion depends crucially on
the combination of impatience and stationarity. While in our model,
there are no such assumptions since in Model 1 the insider receives
the information by one time and there exist no channels such like
time discounting, the public revelation and the obsolescence of
private information that generate cost linked to impatience. Last
but not least, compared to  Kyle (1985), risk-averse insider in
Model 1 obtains greater guaranteed profits at the cost of smaller
risky profits estimated at the beginning of trading.

 Model 2 shows that the risk-neutral insider
 transfers her information to public price gradually when trading happens
frequently. In discrete time case, by producing a more profitable
market depth, the insider is able to obtain a larger ex ante
expectation of total profits across all periods than that in Kyle
(1985). In the limit as the number of trading periods becomes
infinity, however, the insider cannot be rewarded by additional
ability of  affecting the pricing rule. In fact, the difference in
equilibriums between Model 2 and Kyle (1985) is disappearing as
trading frequency is growing, since that in limit the constant
liquidity parameter given by market maker in Kyle (1985) is exactly
the one the insider would like to choose provided her with such
discretion.

In Model 3, the risk-seeking insider has an incentive to postpone
trades to the future to create greater risk in future profits. Thus,
strategic trading in Model 3 is in sharp contrast with those in
Model 1, 2 and Kyle (1985)  in that, insider prefers to the trading
pattern with less information released early on and greater
information revealed latter to keep information advantage as long as
possible. Moreover, Model 3 allows an increasing liquidity parameter
through trades, consistent with the results motivated by the ``rat
race" effect in Foster and Viswanathan (1996) with competitive
insiders endowed with negatively correlated information.

There are large numbers of literature extending Kyle's (1985)
strategic trading model. Holden and Subrahmanyam (1992) consider the
competition among multiple insiders each endowed with perfect
private information. While Foster and Viswanathan (1996) study the
competition with heterogenous private signals. Huddart Hughes and
Levine (2001) examine the case where insider must announce her
trading volume after the submission while Huddart and Hughes (2004)
study the case with pre-announcement of insider trade. Recently,
Caldentey and Stacchetti (2010) study the extended Kyle model with
insider observing a signal that tracts the evolution of asset's
fundamental value and with a random public announcement time
revealing the current value of asset. A common characterization is
that they all inherit the Kyle assumption that the risk-neutral
insider considers liquidity parameter as a constant given by market
maker. Although several papers such like Holden and Subrahmanyam
(1994), Baruch (2002) and Zhang (2004) extend Kyle model to
accomodate risk-averse insider, still they remain the constant
pricing rule assumption unchanged. Hence, our new improvements on
Kyle model by considering both the possibility of insider's effect
on pricing rule and of the insider's risk different attitudes might
have potential applications on various models based on Kyle (1985).

 The rest of  paper is organized as follows: Section 2 presents our
three models based on analysis of Kyle (1985). Section 3
 focuses on Model 1, deriving the sequential auction equilibrium in
 discrete time setting, showing the limit results when the trading period
 number goes to infinity, and illustrating the endogenous
 parameters numerically. Section 4 and section 5 are devoted to
 Model 2 and Model 3 respectively. Finally, section 6 makes some
 concluding comments.
\begin{center}\end{center}
\begin{center}\section{\small{ ANALYSIS ABOUT  KYLE (1985)
MODEL AND PRESENTATION OF OUR MODELS}}\end{center}

\subsection {Basic Notations}

We conform to the notation of Kyle (1985). A risky asset has a
liquidation value $v$, normally distributed with mean $p_{0}$ and
variance $\sigma^2_{v}$. The asset is traded in $N$ sequential
moments $\{t_{n}\}_{n=1, \cdots, N}$ with $t_{n}=n\Delta t_{N}$ and
$\Delta t_{N}=1/N$. The market participants are insider, market
maker and noise traders. The insider knows the true value $v$ and
she submits trading volume $x_{n}$ in the $nth$ period, with her
profits in the $nth$ period denoted by $\pi_{n}$.  Noise traders'
total demand in the $nth$ period denoted by $u_{n}$ is
exogenously-generated, normally distributed with mean 0 and variance
$\sigma^2_{u}\Delta{t_{N}}$ in the N periods model. Market maker
observes the total trading volume $y_{n}=x_{n}+u_{n}$ prior to the
$nth$ auction, and then absorbs it at price $p_{n}$. An important
assumption following Kyle (1985) in our paper is that $p_{n}$
satisfies

\emph{Assumption 1. (Semi-strong) Market Efficiency:}
\begin{align}\label{eq.2.1}
p_{n}=E[v|y_{1}, y_{2}, \cdots, y_{n}]\qquad for\ n=1, 2, ..., N.
\end{align}

Before presenting our new models, we analyze the roles played by
assumption 1 and the ``constant pricing rule" assumption in Kyle
(1985) .

\subsection { The Constant Pricing Rule Assumption  in Kyle (1985) }

To find an equilibrium, Kyle gives three assumptions -market
efficiency, profit maximization and ``constant pricing rule"
assumption in definition ``linear equilibrium" (page 1321).  We
examine the last assumption carefully and attempt to find out how it
works in searching of the equilibrium.
 Before fixing the optimal value for strategy $x_{n}$ at the $nth$
 period, Kyle claims, in a linear equilibrium, $p_{n}$ is given by
\begin{align}\label{eq.2.2}
p_{n}=p_{n-1}+\lambda_{n} (x_{n}+u_{n}) +h,
\end{align}
where $h$ is some linear function of $x_{1}+u_{1}, \cdots,
x_{n-1}+u_{n-1}$ (page 1324). And then, $\lambda_{n}$ is regarded as
a constant independent with insider's strategy $x_{n}$ in Kyle's
following deduction:
\begin{align*}
&\max\limits_{x_{n}}\{ (v-p_{n-1}-\lambda_{n}x_{n}-h)
x_{n}+\alpha_{n} (v-p_{n-1}
-\lambda_{n}x_{n}-h) ^2+\alpha_{n}\lambda^2_{n}\sigma^2_{u}\Delta{t_{n}}+\delta_{n\}}\\
&\Rightarrow\\
&x_{n}=\frac{1-2\alpha_{n}\lambda_{n}}{2\lambda_{n}
(1-\alpha_{n}\lambda_{n}) } (v-p_{n-1}-h).
\end{align*}
Clearly, by assuming the constant pricing rule in  definition
``linear equilibrium", the insider treats  pricing rule
$\lambda_{n}$ as  unrelated to  her choice.

\subsection { The Strategy Space  in Kyle (1985) }

It appears that in Kyle's model, insider's strategy is chosen from
the space consists of functions measurable to information available
to him. However, careful examination shows that a good property
about $x_{n}$ is actually used before its value being fixed. In
fact,
 (\ref{eq.2.2}) can hold only in the following deduction:
\begin{equation*}
\begin{aligned}
p_{n}&=p_{n-1}+p_{n}-p_{n-1}\\
&=p_{n-1}+E[v-p_{n-1}|y_{1}, y_{2}, \cdots, y_{n-1}, y_{n}] && \text{\mbox{ (by  ``market efficiency" assumption}) }\\
&=p_{n-1}+\lambda_{n} y_{n}+h (y_{1}, \cdots, y_{n-1}) &&
\text{\mbox{ (by normality of $x_{1}, ..., x_{n}$).}}
\end{aligned}
\end{equation*}
If $x_{1}, \cdots, x_{n}$ are not gaussian, the expression of
$p_{n}$ (\ref{eq.2.2}) cannot be certainly ensured.\footnote{ Even
if we suppose $x_{1}, \cdots, x_{n}\in L^2 (\Omega,
\mathfrak{F}_{n}, P ) $ with $\mathfrak{F}_{n}=\sigma\{x_{1}+u_{1},
\cdots, x_{n}+u_{n}, v\}$, (\ref{eq.2.2}) needs not hold. Generally,
by definition of conditional expectation
 (Kallenberg, Olav (2002), page 103, 104), $E[v|x_{1}+u_{1}, \cdots,
x_{n}+u_{n} ]$ is just an orthogonal Hilbert space projection of $v$
onto the linear subspace $L^2 (\Omega, \mathfrak{F}_{n}, P ) $.} The
conditional expectation in the above, usually as a nonlinear
measurable function of $x_{1}+u_{1}, \cdots, x_{n}+u_{n}$, is hard
to get an explicit expression. Subsequently,  the cumulative profits
(ie., Eq.(3.24) in page 1324) satisfying
\begin{align*}
&E[\sum^N\limits_{k=n}\pi_{k}|p_{1}, p_{2}, \cdots, p_{n-1}, v]\\
&= (v-E (v|x_{1}+u_{1}, \cdots, x_{n}+u_{n})) +\alpha_{n} (v-E
(v|x_{1}+u_{1}, \cdots, x_{n}+u_{n})) ^2+\delta_{n}
\end{align*}
is not an explicit function of $x_{n}$ any more and thus the
maximization problem cannot be solved. Eventually, all relationships
that build upon the method of backward induction will no longer
hold. In conclusion, insider's strategy in Kyle (1985) model is
actually chosen from the gaussian space.

\subsection { An Inconsistency Implied in Kyle model}

Given gaussian strategy in each period, we know the orders $y_{1},
y_{2}, \cdots, y_{n}$ are normally distributed variables. The
orthogonalization of $y_{1}, y_{2}, \cdots, y_{n}$ produces:

$$\widetilde{y}_{1}, \widetilde{y}_{2}, \cdots, \widetilde{y}_{n}$$
in which $\widetilde{y}_{i}=y_{i}-\sum^{i-1}\limits_{k=1}\frac{cov
(y_{i}, y_{k}) }{cov (y_{k}, y_{k}) }y_{k}$ represents the surprise
in the $ith$ $ (1\leq i\leq n) $ total trading volume. The
assumption of market efficiency (\ref{eq.2.1}) implies
\begin{equation*}
\begin{aligned}
p_{n}-p_{n-1}=&E (v-p_{n-1}|y_{1}, \cdots, y_{n}) =E
 (v-p_{n-1}|\widetilde{y_{1}}, \cdots, \widetilde{y_{n}}) =E
 (v-p_{n-1}|\widetilde{y_{n}}).
\end{aligned}
\end{equation*}
Thus
\begin{align}{\label{eq.2.3}}
p_{n}-p_{n-1}=\lambda_{n}\widetilde{y_{n}}\quad with \quad
\lambda_{n}=\frac{\beta_{n}\Sigma_{n-1}}{{\beta_{n}}^2\Sigma_{n-1}+\sigma^{2}_{u}\Delta{t_{N}}}.\end{align}
Obviously, the informed submission  $x_{n}$ does affect the pricing
rule through the  trading intensity $\beta_{n}$. Thus, an
inconsistency yields between the implications of the ``market
efficiency" assumption and  of the ``constant pricing rule".

 In a semi-strong efficient market, insider with the gaussian
 strategy (thus, (\ref{eq.2.3}) holds.) has an incentive to deviate
 from the ``optimal strategy" depicted in the equilibrium of Kyle
 (1985) to create a more profitable pricing rule. To maintain the
 market efficiency, the market maker would adjust the price
 according to the new strategy insider will choose.
 Interestingly, for Kyle's one period model, equilibrium is the
 same whether or not the
 insider ignores the effect her strategy on pricing rule.
Admati and Pfleiderer (1998) also notice this virtue possessed by
the one period Kyle model. This coincidence, when $N=1$, is
equivalent to the fact that Kyle's equilibrium satisfies
\begin{align}\label{eq.2.4}
\lambda^{'}_{1} (\beta_{1}) =0.\end{align} However, for a general
period number such as $N=2$, (\ref{eq.2.4}) does not hold any more.

\begin{proposition}\label{pro.1}
In equilibrium of the two periods Kyle (1985) model, the pricing
rule in the first period satisfies
\begin{align}\label{eq.2.5}
\lambda^{'}_{1} (\beta_{1}) > 0.
\end{align}
\end{proposition}
Proposition \ref{pro.1} shows that in the first period of the two
periods model, those informed submissions around the optimum in Kyle
(1985) have positive effect on the pricing rule.

\subsection {Presentation of Our Models}

Note that $\sigma\{y_{1}, y_{2}, \cdots, y_{n-1}, v\}=\sigma\{y_{1},
y_{2}, \cdots, y_{n-1}, v-p_{n-1}\}$ since the price $p_{n-1}$,
satisfying (\ref{eq.2.1}), is measurable to the historical
information $y_{1}, y_{2}, \cdots, y_{n-1}$. Thus, each gaussian
strategy $x_{n}$ measurable to $ \sigma\{y_{1}, y_{2}, \cdots,
y_{n-1}, v\}$ has the following form:
\begin{align}\label{eq.2.6}x_{n}=\beta_{n} (v-p_{n-1}) +b_{n} (y_{1}, \cdots, y_{n-1})
+c_{n}\end{align} in which $\beta_{n}, c_{n}\in  \mathbb{R}$ and
$b_{n} (y_{1}, \cdots, y_{n-1}) $ is a linear function of $y_{1},
\cdots, y_{n-1}$. Under assumption (2.1), the following proposition
characterizes the profits of insider with any submission $
(\ref{eq.2.6}) $.

\begin{proposition}\label{pro.2}
Under assumption (2.1), when insider adopts strategies such as
 (\ref{eq.2.6}), there exist non-random real numbers $\alpha_{n-1}$,
$h_{n-1}$, $\delta_{n-1}$, $n=1, 2, \cdots, N+1$, such that
\begin{align}\label{eq.2.7}
E (\sum^{N}_{k=n}\pi_{k}|p_{1}, p_{2}, \cdots, p_{n-1}, v)
=\alpha_{n-1} (v-p_{n-1}) ^2+h_{n-1} (v-p_{n-1}) +\delta_{n-1},
\end{align}
where \begin{align}\label{eq.2.8} &\alpha_{n-1}=\alpha_{n}
 (1-\lambda_{n}\beta_{n}) ^2+\beta_{n} (1-\lambda_{n}\beta_{n}), \\
\label{eq.2.9} &h_{n-1}= (b_{n} (y_{1}, \cdots, y_{n-1})
+c_{n}+h_{n})
 (1-\lambda_{n}\beta_{n}), \\
\label{eq.2.10}
&\delta_{n-1}=\delta_{n}+\alpha_{n}\lambda^2_{n}\sigma^{2}_{u}\Delta{t_{N}}.
\end{align}
with $\alpha_{N}=h_{N}=\delta_{N}=0$ and $\lambda_{n}$ satisfying
 (\ref{eq.2.3}).
 In (\ref{eq.2.6}), the  submission structured on historical
information $b_{n} (y_{1}, \cdots, y_{n-1}) $ and insider's average
submission  $c_{n}$ can  affect the conditional profits
 (\ref{eq.2.7}) only through the term $h_{n-1} (v-p_{n-1}) $. Moreover,
$b_{n} (y_{1}, \cdots, y_{n-1}) $ and $c_{n}$ cannot affect any of
$\Sigma_{k}, \lambda_{k}, $ or $p_{k}$ with $1\leq k\leq N$.
\end{proposition}

Proposition \ref{pro.2} shows that, in the nth period, the
submission structured on common knowledge $b_{n} (y_{1}, y_{2},
\cdots, y_{n-1}) +c_{n}$ yields zero profits in ex ante expectation.
Moreover, when insider chooses  $b_{n} (y_{1}, y_{2}, \cdots,
y_{n-1}) +c_{n}$ unbounded and  the other parameters bounded in her
plan, then at the case $v-p_{n-1}>0 (<0) $, her conditional profits
(loss) will be unbounded.
 To avoid the technical trouble with thus strategies that always yield
zero profits in ex ante expectation and can
 yield infinite profits in absolute value, we consider a limited strategy space on insider. That is,
the space of strategies constructed on the estimation error
$v-p_{n-1}$ which represents the totally unrevealed information in
the sense it is independent with historical information $y_{1},
y_{2}, \cdots, y_{n-1}$ and thus exclusively known to the insider:
\begin{align}\label{eq.2.11}X_{n}=\{\beta_{n} (v-p_{n-1}) |\beta_{n}\in \mathbb{R}\}.\end{align}
 (\ref{eq.2.11}) is actually the strategy space that contains the
``optimal" informed strategy depicted in Kyle(1985). Note that
$x_{n}\in X_{n}$ implies $\widetilde{y_{n}}=y_{n}$ in
(\ref{eq.2.3}), and this will be used throughout the rest of the
article.

As shown by (\ref{eq.2.7}) and (\ref{eq.2.9}), insider with strategy
$x_{n}\in X_{n}$ acquires profits accumulated from the nth period to
the end:
\begin{align}\label{eq.2.12}
E (\sum^{N}_{k=n}\pi_{k}|p_{1}, p_{2}, \cdots, p_{n-1}, v)
 =\alpha_{n-1} (v-p_{n-1})
^2+ \delta_{n-1}.
\end{align}
The optimal strategy $\beta_{n} (v-p_{n-1}) $ , or equivalently,
$\beta_{n}$, should be determined in equilibrium by
profit-maximization principle. However, generally, (\ref{eq.2.12})
has no maximization due to the partial ordering among conditional
expectations and thus directly maximizing
 (\ref{eq.2.12}) is meaningless.
 Note that
insider's conditional profits consist of two different terms, the
risky profits $\alpha_{n-1} (v-p_{n-1}) ^2$ as the source of risk,
and guaranteed profits $\delta_{n-1}$ that cannot be affected by the
realization of $v$ or $p_{n-1}$. Naturally, three  models, each with
the Assumption 1 and a different profit-maximization principle
stated by Assumption 2 are presented.

\emph{ Assumption 2. Profit Maximization: At the nth ($1\leq n \leq
N$) period, with the informed strategy space (\ref{eq.2.11}) and the
informed profits (\ref{eq.2.12}),
\begin{itemize}
\item Model 1 (the risk-averse insider model):
the insider firstly maximizes the guaranteed profit
$$\max\limits_{\beta_{n}}\delta_{n-1}, $$ and secondly she maximizes
the risky profits $$\max\limits_{\beta_{n}\in\{argmax{\
                                       \delta_{n-1}}\}}\alpha_{n-1} (v-p_{n-1}) ^2.$$
\item  Model 2 (the risk-neutral insider model):
the insider  maximizes the ex ante expectation
$$\max\limits_{\beta_{n}}E (\sum^{N}_{k=n}\pi_{k}). $$
\item Model 3 (the risk-seeking insider model):
the insider firstly maximizes the risky profit
$$\max\limits_{\beta_{n}}\alpha_{n-1} (v-p_{n-1}) ^2, $$ and
secondly she maximizes the guaranteed profit
$$\max\limits_{\beta_{n}\in\{argmax{\   \alpha_{n-1} (v-p_{n-1}) ^2}\}}\delta_{n-1}.$$
\end{itemize}}

\begin{center} \section{\small{MODEL 1: THE EQUILIBRIUM OF THE RISK-AVERSE INSIDER
MODEL}} \end{center}  \setcounter{equation}{0}

\subsection {The Discrete Equilibrium}

Theorem 1 characterizes a sequential auction equilibrium with
endogenous parameters expressed by a difference equation system.

\begin{theorem}\label{theorem.1}
In Model 1 with  trading period number $N$, a subgame perfect linear
equilibrium exists. In this equilibrium, there are real numbers
$\beta_{n}, \lambda_{n}, \alpha_{n}$ and $\Sigma_{n}$, such that
\begin{align}\label{eq.3.1}
&x_{n}=\beta_{n} (v-p_{n-1}), \\
\label{eq.3.2}
&p_{n}-p_{n-1}=\lambda_{n}y_{n},\\
\label{eq.3.3}
&\Sigma_{n}=var (v|y_{1}, y_{2}, \cdots, y_{n}), \\
\label{eq.3.4} &E (\sum^{N}_{k=n}\pi_{k}|p_{1}, p_{2}, \cdots,
p_{n-1}, v) =\alpha_{n-1} (v-p_{n-1}) ^2+\delta_{n-1}.
\end{align}
The above  real numbers  $\beta_{n}, \alpha_{n}$ and $\Sigma_{n}$
can be represented as:
\begin{align}\label{eq.3.5}
&\delta_{n}=a_{n}\sigma_{u}{\Delta{t_{N}}}^{1/2}{\Sigma_{n}}^{1/2}&
(n=0, 1, 2, \cdots, N-1),&\\
\label{eq.3.6}
&\alpha_{n}=b_{n}\sigma_{u}{\Delta{t_{N}}}^{1/2}{\Sigma_{n}}^{-1/2}
 &\quad (n=0, 1, 2, \cdots, N-1),&\\
\label{eq.3.7}
&\beta_{n}=c_{n}\sigma_{u}{\Delta{t_{N}}}^{1/2}{\Sigma_{n-1}}^{-1/2}
& \quad (n=1, 2, \cdots, N).&
\end{align}
 in which the sequences $\{a_{n}\}$, $\{b_{n}\}$, $\{c_{n}\}$, subject to
terminal values $a_{N-1}=0$, $b_{N-1}=\frac{1}{2}$, $c_{N}=1$, are
given recursively:
\begin{align}\label{eq.3.8}
&a_{n-1}=a_{n} (\frac{1}{c^2_{n}+1}) ^{1/2}+b_{n} (\frac{1}{c^2_{n}+1}) ^{3/2}c^2_{n},\\
\label{eq.3.9}
&b_{n-1}=b_{n} (\frac{1}{c^2_{n}+1}) ^{3/2}+\frac{c_{n}}{c^2_{n}+1},\\
\label{eq.3.10} &c_{n}= (\frac{2b_{n}-a_{n}}{a_{n}+b_{n}}) ^{1/2},
\end{align}
where $c_{n}>0, n=1, 2, \cdots, N$.
\end{theorem}
\emph{PROOF:}
 The proof is by backward induction. The problem in
the last period is choosing the optimal strategy $x_{N}=\beta_{N}
 (v-p_{N-1}) $ or equivalently, $\beta_{N}$, in the maximization problem:
\begin{equation}{\label{eq.3.11}}
\begin{aligned}
\max\limits_{\beta_{N}}E[\pi_{N}|p_{1}, \cdots, p_{N-1}, v]
&=\max\limits_{\beta_{N}}E[x_{N} (v-p_{N}) |y_{1}, \cdots, y_{N-1}, v]\\
&=\max\limits_{\beta_{N}}E[x_{N} (v-p_{N-1}-\lambda_{N}y_{N}) |y_{1}, \cdots, y_{N-1}, v]\\
&=\max\limits_{\beta_{N}}\beta_{N} (1-\lambda_{N}\beta_{N}) (v-p_{N-1}) ^2\\
&=\max\limits_{\beta_{N}}\frac{\beta_{N}\sigma^2_{u}\Delta{t_{N}}}{\beta^{2}_{N}\Sigma_{N-1}+\sigma^2_{u}\Delta{t_{N}}}
 (v-p_{N-1}) ^2.
\end{aligned}
\end{equation}
As seen from (\ref{eq.3.11}), with the amount of information
available $\Sigma^{1/2}_{N-1}$ and the estimation error $v-p_{N-1}$,
the insider will choose the maximizing value
$$\beta_{N}=\sqrt{\frac{\sigma^2_{u}\Delta{t_{N}}}{\Sigma_{N-1}}}.$$
Thus, (\ref{eq.3.4}) holds when $n=N$ with
\begin{align*}
\alpha_{N-1}=\frac{1}{2}\sqrt{\frac{\sigma^2_{u}\Delta{t_{N}}}{\Sigma_{N-1}}},
\quad \delta_{N-1}=0.
\end{align*}
In general, if in the $n+1th$ period, insider's optimal strategy
$\beta_{n+1}$ supports profits
\begin{align*}
E[\sum^N_{k=n+1}\pi_{k}|p_{1}, \cdots, p_{n}, v]=\alpha_{n}
 (v-p_{n}) ^2+\delta_{n}
\end{align*}
with\begin{align}\label{eq.3.12}
 \delta_{n}=a_{n}\sigma_{u}
\Delta t_{N}^{1/2} \Sigma_{n}^{1/2}, \ \alpha_{n}=b_{n}\sigma_{u}
\Delta t_{N}^{1/2} \Sigma_{n}^{-1/2}, \
\beta_{n+1}=c_{n+1}\sigma_{u} \Delta t_{N}^{1/2} \Sigma_{n}^{-1/2}.
\end{align}
Then, any informed submission $x_{n}=\beta_{n} (v-p_{n-1}) $ can
expect to yield
\begin{equation}{\label{eq.3.13}}
\begin{aligned}
E[\sum^N_{k=n}\pi_{k}|p_{1}, \cdots, p_{n-1}, v]
&=E[\sum^N_{k=n+1}\pi_{k}+x_{n} (v-p_{n}) |y_{1}, \cdots, y_{n-1}, v]\\
&=\alpha_{n-1} (v-p_{n-1}) ^2+\delta_{n-1}
\end{aligned}
\end{equation}
with
\begin{align}\label{eq.3.14}
\alpha_{n-1}&=\alpha_{n} (1-\lambda_{n}\beta_{n}) ^2+\beta_{n} (1-\lambda_{n}\beta_{n}), \\
\label{eq.3.15}
\delta_{n-1}&=\delta_{n}+\alpha_{n}\lambda^2_{n}\sigma^2_{u}\Delta{t_{N}}.
\end{align}
Additionally, by definitions, we have
\begin{align}\label{eq.3.16}
\lambda_{n}&=\frac{\beta_{n}\Sigma_{n-1}}{\beta^2_{n}\Sigma_{n-1}+\sigma^2_{u}\Delta{t_{N}}},\\
\label{eq.3.17}
\Sigma_{n}&=\frac{\Sigma_{n-1}\sigma^2_{u}\Delta{t_{N}}}{\beta^2_{n}\Sigma_{n-1}+\sigma^2_{u}\Delta{t_{N}}}.
\end{align}
Further, another relationship follows from (\ref{eq.3.16}) and
 (\ref{eq.3.17}):
\begin{align}\label{eq.3.18}
\Sigma_{n}= (1-\lambda_{n}\beta_{n}) \Sigma_{n-1}.
\end{align}
Substituting (\ref{eq.3.12}) (\ref{eq.3.18}) and (\ref{eq.3.16})
into (\ref{eq.3.14}) and into (\ref{eq.3.15}) respectively yields
$\alpha_{n-1}$ and $\delta_{n-1}$ in expression of $\beta_{n}$
\begin{equation}\label{eq.3.19}
\begin{aligned}
\alpha_{n-1}(\beta_{n})
&=b_{n}\sigma_{u}{\Delta{t}}^{1/2}_{N}\Sigma^{-1/2}_{n}
 (1-\lambda_{n}\beta_{n}) ^2+\beta_{n} (1-\lambda_{n}\beta_{n}) && \text{by (\ref{eq.3.14}) and (\ref{eq.3.12}), }\\
&=b_{n}\sigma_{u}{\Delta{t}}^{1/2}_{N}\Sigma^{-1/2}_{n-1}
 (\frac{\sigma^2_{u}\Delta{t_{N}}}{\beta^2_{n}\Sigma_{n-1}+\sigma^2_{u}\Delta{t_{N}}})
^{3/2}+\frac{\beta_{n}\sigma^2_{u}\Delta{t_{N}}}{\beta^2_{n}\Sigma_{n-1}+\sigma^2_{u}\Delta{t_{N}}}&&
\text{by (\ref{eq.3.18}) and (\ref{eq.3.16}) },
\end{aligned}
\end{equation}

\begin{equation}\label{eq.3.20}
\begin{aligned}
\delta_{n-1}(\beta_{n})&=a_{n}\sigma_{u}{\Delta{t}}^{1/2}_{N}
\Sigma^{1/2}_{n}
+b_{n}\sigma_{u}{\Delta{t}}^{1/2}_{N} \Sigma^{-1/2}_{n}{\lambda_{n}^2}\sigma^2_{u}\Delta{t_{N}}&& \text{by (\ref{eq.3.14}) and (\ref{eq.3.12}), }\\
&=a_{n}\sigma^2_{u}\Delta{t_{N}}
 (\frac{\Sigma_{n-1}}{\beta^2_{n}\Sigma_{n-1}+\sigma^2_{u}\Delta{t_{N}}})
^{1/2}
+b_{n} \beta^2_{n}\sigma^2_{u}\Delta{t_{N}} (\frac{\Sigma_{n-1}}{\beta^2_{n}\Sigma_{n-1}+\sigma^2_{u}\Delta{t_{N}}}) ^{3/2}&& \text{by (\ref{eq.3.18}) and (\ref{eq.3.16}) }.\\
\end{aligned}
\end{equation}
Therefore, solving $\max\limits_{\beta_{n}}\delta_{n-1}(\beta_{n})$
yields: when $2b_{n}-a_{n}>0, a_{n}\geq 0, b_{n}\geq 0$ and
$a_{n}+b_{n}>0$
 (since calculation shows the FOC and SOC require
$c_{n} (a_{n}+b_{n}) >0$),
\begin{align*}
\beta_{n}=\pm\sqrt{\frac{2b_{n}-a_{n}}{a_{n}+b_{n}}}\frac{\sigma_{u}{\Delta{t}}^{1/2}_{N}}
{\Sigma^{1/2}_{n-1}}.\end{align*} Due to the fact that
\begin{align*}\alpha_{n-1}
 (\sqrt{\frac{2b_{n}-a_{n}}{a_{n}+b_{n}}}\frac{\sigma_{u}{\Delta{t}}^{1/2}_{N}}
{\Sigma^{1/2}_{n-1}}) >\alpha_{n-1}
 (-\sqrt{\frac{2b_{n}-a_{n}}{a_{n}+b_{n}}}\frac{\sigma_{u}{\Delta{t}}^{1/2}_{N}}
{\Sigma^{1/2}_{n-1}}),\end{align*} the negative $\beta_{n}$ is
excluded. Thus, for $n$,
 (\ref{eq.3.7}) with (\ref{eq.3.10}) holds.
Moreover, substituting (\ref{eq.3.7}) into (\ref{eq.3.19}), we find
$\alpha_{n-1}=b_{n-1}\sigma_{u}{\Delta{t}}^{1/2}_{N}\Sigma_{n-1}^{-1/2}$
with $b_{n-1}$ satisfying (\ref{eq.3.9}). While substituting
(\ref{eq.3.7}) into (\ref{eq.3.20}) yields
$\delta_{n-1}=a_{n-1}\sigma_{u}{\Delta{t}}^{1/2}_{N}\Sigma_{n-1}^{1/2}$
with $a_{n-1}$ satisfying (\ref{eq.3.8}).

The difference equation system (\ref{eq.3.8}) (\ref{eq.3.9}) and
 (\ref{eq.3.10}) gives a unique solution for sequences $\{a_{n}\}, \{b_{n}\}$ and $\{c_{n}\}$
with terminal value $a_{N-1}, b_{N-1}$ and $c_{N}$. In fact, if
$a_{n}, b_{n}$ and $c_{n+1}$ are fixed, then (\ref{eq.3.10}) yields
$c_{n}$, and in turn, substituting $c_{n}$ into (\ref{eq.3.8})
yields $a_{n-1}$, and into (\ref{eq.3.9}) yields $b_{n-1}$.

At last, verify the conditions $2b_{n}-a_{n}>0, a_{n}\geq 0,
b_{n}\geq 0$ and $a_{n}+b_{n}>0$ as follows. In the last periods,
these conditions hold since $a_{N-1}=0, b_{N-1}=\frac{1}{2}$.
Suppose for a general $n$, theses conditions hold, then from the
above proof, (\ref{eq.3.8}), (\ref{eq.3.9}) and
 (\ref{eq.3.10}) with $c_{n}> 0$ can be acquired and substituting
them into $2b_{n-1}-a_{n-1}$ yields
\begin{align*}
2b_{n-1}-a_{n-1} = (\frac{1}{c^2_{n}+1}) ^{3/2}[
 (2b_{n}-a_{n} (c^2_{n}+1)) +2c_{n} (c^2_{n}+1)
^{1/2}-b_{n}c^2_{n}]=\frac{2c_{n}}{c^2_{n}+1} >0.
\end{align*}
Further,
\begin{align*}b_{n-1}=b_{n} (\frac{1}{c^2_{n}+1})
^{3/2}+\frac{c_{n}}{c^2_{n}+1}>0,\ a_{n-1} =a_{n}
 (\frac{1}{c^2_{n}+1}) ^{1/2}+b_{n} (\frac{1}{c^2_{n}+1})
^{3/2}c^2_{n}>0,\end{align*} and $a_{n-1}+b_{n-1}>0$ is obvious.
 In conclusion, these
conditions
hold and hence Theorem 1 is proved.  $ \hfill Q.E.D.$\\

Recall that in Kyle (1985), insider's optimal submission depends the
value of $\lambda_{n}$ announced by market maker. In equilibrium of
Model 1, however, it is expressed explicitly as proportional to the
product of amount of camouflage and inverse of information available
$\sigma_{u}\Delta{t^{1/2}_{N}}\Sigma^{-1/2}_{n-1}$, involving no
pricing rule $\lambda_{n}$.
 On the other
hand, like Kyle (1985), inspecting insider's profits shows that both
the expectation of risky profits $\alpha_{n-1} (v-p_{n-1}) ^2$ and
guaranteed profits $\delta_{n-1}$ are proportional to the product of
amount of camouflage and information available
$\sigma_{u}\Delta{t^{1/2}_{N}}\Sigma^{1/2}_{n-1}$. And inspecting
the other parameters shows, if $\sigma_{u}$ doubles, then
$\alpha_{n}, \beta_{n}, \delta_{n}$ double, $\Sigma_{n}$ by
(\ref{eq.3.17}) is unaffected and $\lambda_{n}$ by (\ref{eq.3.16})
halves. Additionally, the SOC $c_{n}>0$ rules out the strategies
trading inversely on private information which are also not present
in Kyle (1985).

As seen from (\ref{eq.3.7}), the trading intensity sequence
$\{\beta_{n}\}$ is increasing over nearly all trading rounds for two
reasons. (i) By (\ref{eq.3.17}), the information available
$\Sigma^{1/2}_{n-1}$ is decreasing over trading, and thus the
insider needs to trade more intensively to reveal the private
information when it is of less scale. (ii) As will be shown, $c_{n}$
is increasing with $n$ when $n<N$. This is consistent with the
decreasing concern about the effect current trades has on the future
with time going on. Literatures examining the increasing trading
intensities usually notice the second reason but ignore the first
one, might because trading intensities in Kyle (1985) are expressed
less explicitly than us.

 As to the
relationship between the insider's strategy and her profits,
equation (\ref{eq.3.10}) gives some intuitive insights. A simple
calculation shows, the  insider's trading intensity coefficient
$c_{n}$ is increasing with $b_{n}/a_{n}$ i.e., the ratio of marginal
guaranteed profits to marginal risky profits estimated in the next (
$n+1th$ ) period. Thus, if $a_{n}$ is relatively large, then the
insider has an  incentive to trade less currently to keep more
information for the next period since she has a large ability to
acquire high guaranteed profits in the next period.

A special case can be solved easily is the two periods case.
Firstly, recall that Kyle (1985) model has an equilibrium when N=2
 (see Huddart etal., (2001, Proposition 1)), with endogenous
parameters added an upper index $ (0) $ to distinguish from those in
our models, satisfying
\begin{align}\label{eq.3.210}
\alpha^{ (0) }_{0}\approx 0.7495
{\Sigma_{0}}^{-1/2}\sigma_{u}{\Delta t_{N}}^{1/2}, \qquad \delta^{
(0) }_{0}\approx 0.1281 {\Sigma_{0}}^{1/2}\sigma_{u}{\Delta
t_{N}}^{1/2}.
\end{align}
Whereas in  Model 1 with  endogenous parameters added an upper index
$ (1) $, we have:
\begin{align*}
\alpha^{ (1) }_{0}=[\frac{1}{2} (\frac{1}{3})
^{3/2}+\frac{\sqrt{2}}{3}] {\Sigma_{0}}^{-1/2}\sigma_{u}{\Delta
t_{N}}^{1/2}<\alpha^{ (0) }_{0}, \qquad \delta^{ (1) }_{0}=
(\frac{1}{3}) ^{3/2}{\Sigma_{0}}^{1/2} \sigma_{u}{\Delta
t_{N}}^{1/2}>\delta^{ (0) }_{0}.
\end{align*}
Clearly, compared to Kyle (1985), the insider can expect to obtain
increased guaranteed profits at the cost of decreased risky profits
at the beginning of trading. For a general $N>2$, to examine how the
unrevealed information $\Sigma_{n}$, the liquidity parameter
$\lambda_{n}$  and the trading intensity $\beta_{n}$ perform, we
need to resort to the numerical method, before that, the limit
results when $N\rightarrow\infty$ can give some important
theoretical guidance.

\subsection {The Limit Behavior When $N\rightarrow\infty$}

Denote $[Nt]$  the integral part of $Nt$. Taken sequence $\{c_{n}\}$
for example, let $t>0$, the value
$\lim\limits_{\Delta{t_{N}}\rightarrow 0}c_{[Nt]}$
 corresponds to the continuous version of this sequence at time $t$.
Another kind of limit we are interested in is
$\lim\limits_{\Delta{t_{N}}\rightarrow 0}c_{n}$ with a holden $n$,
characterizing what happens just after the beginning of trading. The
former class of limits is called the ``first class of limits", and
the latter is called the ``second class of limits" in Holden and
subrahmanyam (1992). To start the work, the following proposition
establishes some preliminary results for the  main results.
\begin{proposition}\label{pro.3}
The sequence $\{c_{n}\}$ in Theorem 1 can be achieved as follows.
 Given $c_{n}$ and $c_{n-1}$, $c_{n-2}$ is determined  by
 the unique root in $ (0, 1) $ of equation
\begin{align}\label{eq.3.22}
 (c^2_{n}+1) (2-c^2_{n-2}) c^3_{n-1}=2 (1+c^2_{n-1}) ^{1/2}c_{n}c^2_{n-2}
\end{align}
with terminal values $c_{N-1}=\sqrt{2}, c_{N}=1$. Moreover, the
following monotonicity holds:
\begin{align}\label{eq.3.23}
c_{N-1}>c_{N-2}>\cdots, c_{n}>c_{n-1}>c_{2}>c_{1}.\end{align}
$b_{n}$ has a representation of $c_{n}$ and $c_{n-1}$
\begin{align}\label{eq.3.24}
b_{n}=\frac{c_{n} (c^2_{n}+1) ^{1/2} (2-c^2_{n-1}) }{3c^2_{n-1}},
\end{align}
and also does $a_{n}$
\begin{align}\label{eq.3.25}
a_{n}=\frac{2-c^2_{n}}{c^2_{n}+1}b_{n}.
\end{align}
\end{proposition}

Based on Proposition \ref{pro.3}, we obtain all endogenous
  parameters' limits when trading frequency goes to infinity as well as the speeds of
convergence with which the limits are obtained.\footnote{In fact,
continuous versions for discrete parameters often take values $0$ or
$\infty$, singly providing little for depiction of the realistic
situation with frequent but not continuous trading. While when
combined with the convergence speeds, they can make a closer
estimation for the realistic situation.}

\begin{theorem}\label{theorem.2}
When $N\rightarrow\infty$ (equivalently,
$\Delta{t_{N}}\rightarrow{0}$), the sequences $\{a_{n}\}, \{b_{n}\},
\{c_{n}\}$ in Theorem \ref{theorem.1} have limits:
\begin{align}\label{eq.3.26}
&\lim\limits_{N\rightarrow\infty}c_{[Nt]}=0, \\
\label{eq.3.27}
&\lim\limits_{N\rightarrow\infty}b_{[Nt]}=\infty, \\
\label{eq.3.28} &\lim\limits_{N\rightarrow\infty}a_{[Nt]}=\infty,
\end{align} for any $t\in (0, 1) $.
Moreover
\begin{align}\label{eq.3.29}
&\lim\limits_{N\rightarrow\infty}\frac{c_{[Nt]}}{{\Delta{t_{N}}}^{1/4}}= (\frac{2}{3}\frac{1}{1-t}) ^{1/4}, \\
\label{eq.3.30}
&\lim\limits_{N\rightarrow\infty}b_{[Nt]}{\Delta{t_{N}}}^{1/4}= (\frac{2}{3}) ^{3/4} (1-t) ^{1/4},\\
\label{eq.3.31}
&\lim\limits_{N\rightarrow\infty}a_{[Nt]}{\Delta{t_{N}}}^{1/4}=2*
 (\frac{2}{3}) ^{3/4} (1-t) ^{1/4},
\end{align}
and the limits of $c_{n},b_{n}$ and $a_{n}$ with $n$ holden when
$N\rightarrow\infty$ correspond to $t=0$ respectively in above
results.

 The unrevealed information $\Sigma_{n}$ satisfies, for any
$t\in (0, 1]$,
\begin{align}\label{eq.3.32}\lim\limits_{N\rightarrow\infty}\Sigma_{[Nt]}=0\end{align}
with \begin{align}\label{eq.3.33} \Sigma_{0}exp\{-ln (1+c^2_{N-1})
{\Delta{t_{N}}}^{-1}\}<\Sigma_{[Nt]} <\Sigma_{0}exp\{-ln (1+c^2_{1})
{\Delta{t_{N}}}^{-1}\}.
\end{align}
The liquidity parameter $\lambda_{n}$ satisfies, for a holden $n$,
\begin{align}\label{eq.3.34}\lim\limits_{N\rightarrow\infty}\lambda_{n}=\infty,\end{align}
and for any $t\in (0, 1]$,
\begin{align}\label{eq.3.35}\lim\limits_{N\rightarrow\infty}\lambda_{[Nt]}=0.\end{align}
The trading intensity $\beta_{n}$ satisfies, for a holden $n$,
\begin{align}\label{eq.3.36}\lim\limits_{N\rightarrow\infty}\beta_{n}=0,\end{align} and for any $t\in
 (0, 1]$,
\begin{align}\label{eq.3.37}\lim\limits_{N\rightarrow\infty}\beta_{[Nt]}=\infty.\end{align}
\end{theorem}

Estimated at the beginning of trading, Model 1 in limit implies
infinity guaranteed profits and zero risky profits since Theorem 2
shows that $a_{n}$ and $b_{n}$ with n holden are of order
$\Delta{t^{-1/4}_{N}}$ in (\ref{eq.3.5}) and (\ref{eq.3.6}). Thus
the results in the two periods case are confirmed.

Now, examine the limit behaviors of the endogenous parameters-
unrevealed information, liquidity parameter and trading intensity.
  The first class of limits
of unrevealed information sequence
 (\ref{eq.3.32}) shows that the private information is dissipated
immediately (in an arbitrary small time horizon). Thus the
strong-form efficiency characterized by Fama (1970) can be realized
within any positive time $t$ when trading happens continuously.
Inequality (\ref{eq.3.33}) indicates that when $N$ increases, prices
incorporate the fundamental value at an exponential speed.
(\ref{eq.3.37}) says, the trading intensity goes to infinity at any
calendar time $t>0$. While the second class of limits
(\ref{eq.3.36}) shows that in the initial periods, the insider
trades little on private information. It is striking that the
initial low trading intensities (or expected trading
volumes\footnote{Indeed,
$E\frac{1}{2}|x_{n}|=\frac{1}{\sqrt{2\pi}}\beta_{n}\sqrt{\Sigma_{n-1}}$.
Therefore, when $N$ is large enough, for a holden $n$,
$E\frac{1}{2}|x_{n}|\approx\frac{1}{\sqrt{2\pi}} (\frac{2}{3})
^{1/4} (\Delta t_{N}) ^{3/4}$ and for any $t>0$,
$E\frac{1}{2}|x_{[Nt]}| \approx\frac{1}{\sqrt{2\pi}} (\frac{2}{3
(1-t) }) ^{1/4} (\Delta t_{N}) ^{3/4}$. Thus the initial periods
possess relatively low expected trading volumes.}) can eliminate
most information asymmetry so quickly. This stems from the fact that
the market maker responses to every unit of the order extremely
sensitively in initial periods, as shown by (\ref{eq.3.34}). By
contrast, by (\ref{eq.3.35}), at any positive calendar time $t$ when
market maker has lost the high sensitivity due to the little scale
of unrevealed information, the insider would like to trade extremely
aggressively on her information.

These results are similar to those of Holden and Subrahmanyam (1992)
model with competitive risk-neutral insiders and  are in substantial
contrast to those of the Kyle (1985) with a single insider. Chau and
Vayanos (2008) also establishes a monopolistic insider trading model
in which the market obtains a strong form efficiency when trading
frequency is sufficiently large. In their model, with stationary in
the pattern of information arrival, the insider chooses to trade
quickly to avoid costs linked to impatience generated by
time-discounting, public revelation of information or mean-reverting
profitability. Unlike them, the strong form efficiency in our model
is motivated by the risk aversion of insider. Moreover, recall that
in Chau and Vayanos (2008) (also in Caldentey and Stacchetti (2010)
after an endogenous time) , the insider can earn positive profits
even when $\Delta{t_{N}}\rightarrow0$, since although the market
maker's estimation error goes to zero, the insider can compensate
this by trading infinitely intensively on the new private
information flowing in. In our model, however, the  market maker's
estimation error goes to zero at an exponential speed, larger than
the polynomial speed at which the insider speeds up her trading, and
thus insider's profits vanish very soon.

\subsection {Numerical Results}

To illustrate Model 1 numerically, we consider the model's
implication in a variety of settings. We are interested in how
information is released through trading, how the insider's trading
intensity changes through time, how the market maker adjusts price
in response to the order flow, and how lengthening the number of
trading periods affects some of these parameters.
\begin{figure}
\centering \mbox{ \subfigure[Unrevealed information $\Sigma_{n}$
with changing period numbers $N=5, N=20,
N=100$.]{\includegraphics[width=.47\textwidth]{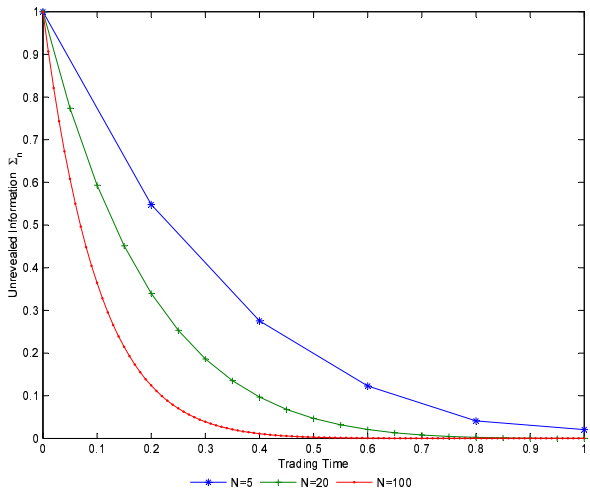}
 }\qquad
 \subfigure[ The approximation by limit results and the actual discrete results when N=1000.]{\includegraphics[width=.47\textwidth]{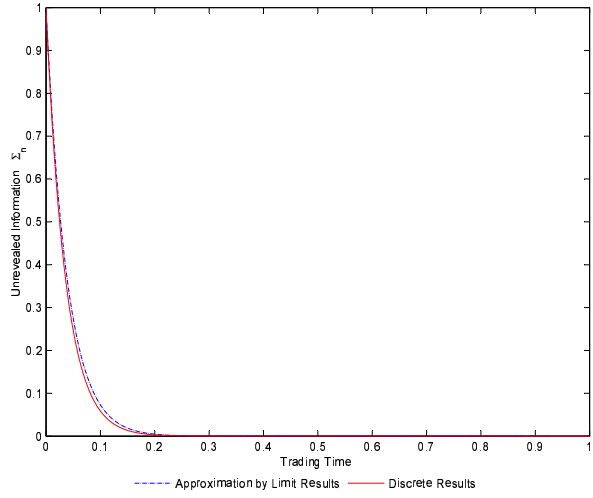}}
} \caption{ (Model 1)  Numeric solutions of the unrevealed
information $\Sigma_{n}$ with one unit of initial variance of
information, half unit of noise trader variance across all periods.}
\end{figure}
\begin{figure}
\centering \mbox{ \subfigure[Liquidity parameter $\lambda_{n}$ with
changing period numbers $N=5, N=20,
N=100$.]{\includegraphics[width=.47\textwidth]{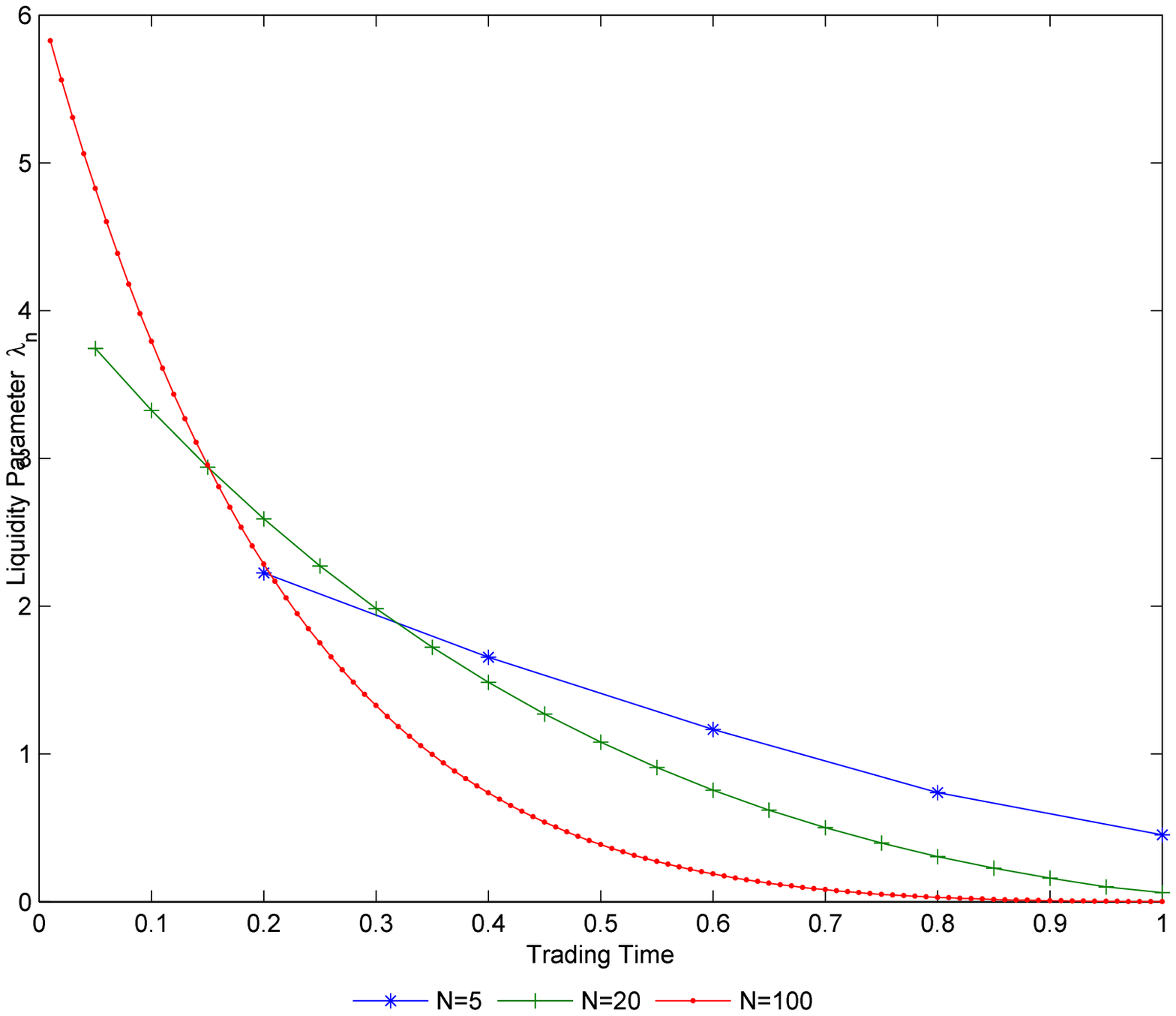}
 }\qquad
 \subfigure[The approximation by limit results and the actual discrete results when N=1000. ]{\includegraphics[width=.47\textwidth]{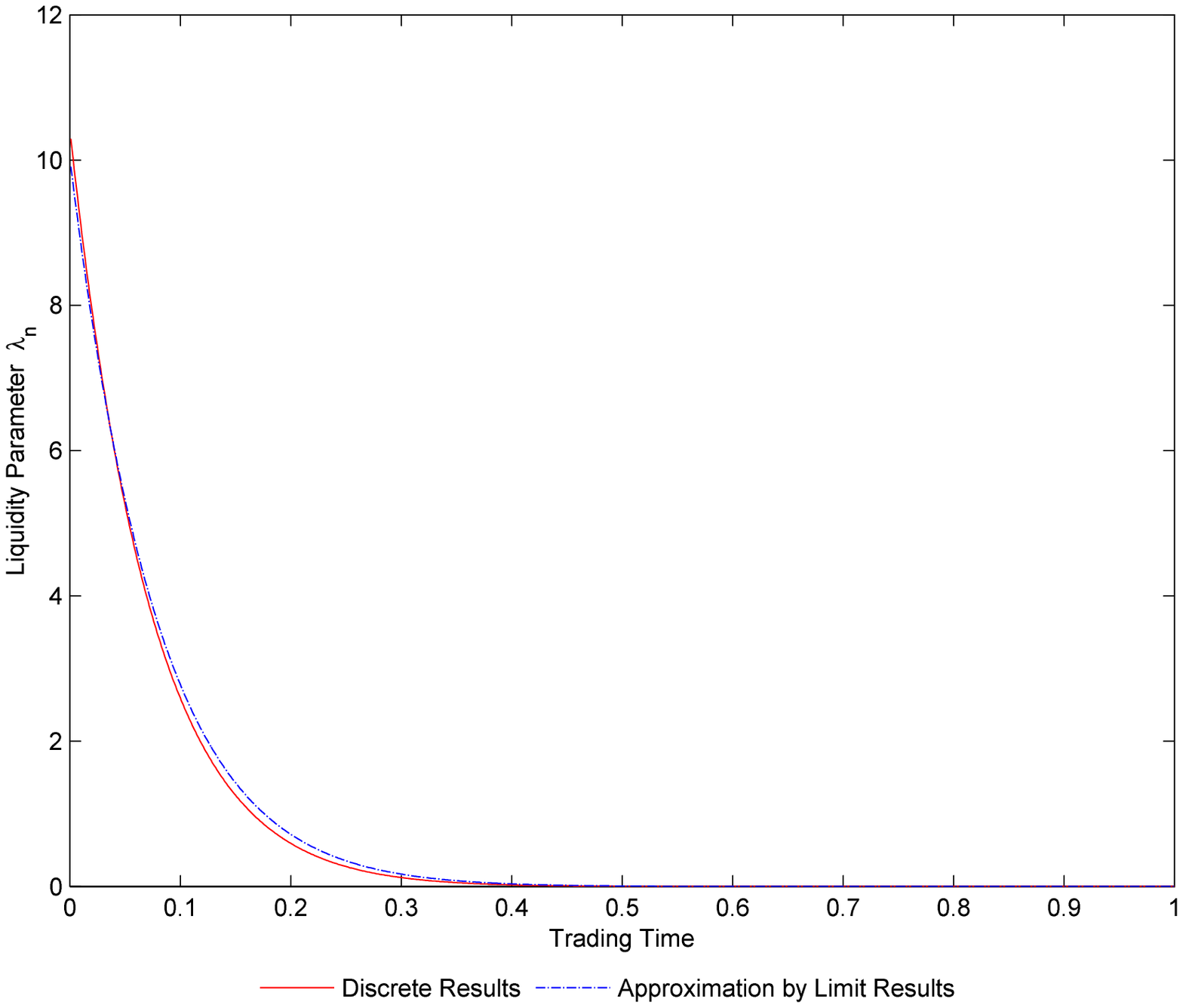}}
} \caption{ (Model 1)  Numeric solutions of the liquidity parameter
$\lambda_{n}$ with one unit of initial variance of information, half
unit of noise trader variance across all periods.}
\end{figure}
\quad
\begin{figure}
\centering \mbox{ \subfigure[Intensity trading on private
information $\beta_{n}$ with changing period numbers $N=5, N=20,
N=100$.]{\includegraphics[width=.47\textwidth]{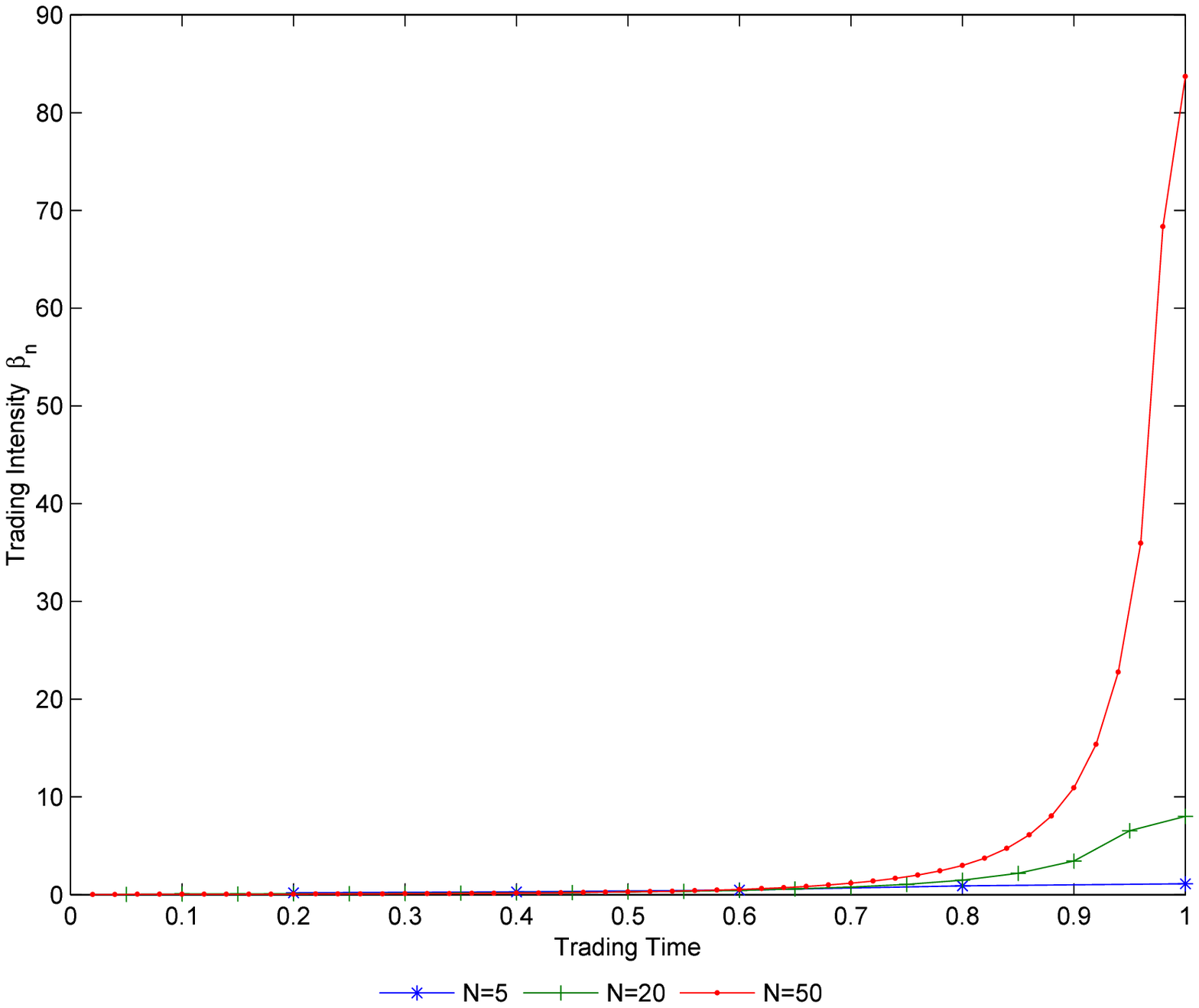}
 }\qquad
 \subfigure[ The approximation by limit results and the actual discrete results when N=50.]{\includegraphics[width=.47\textwidth]{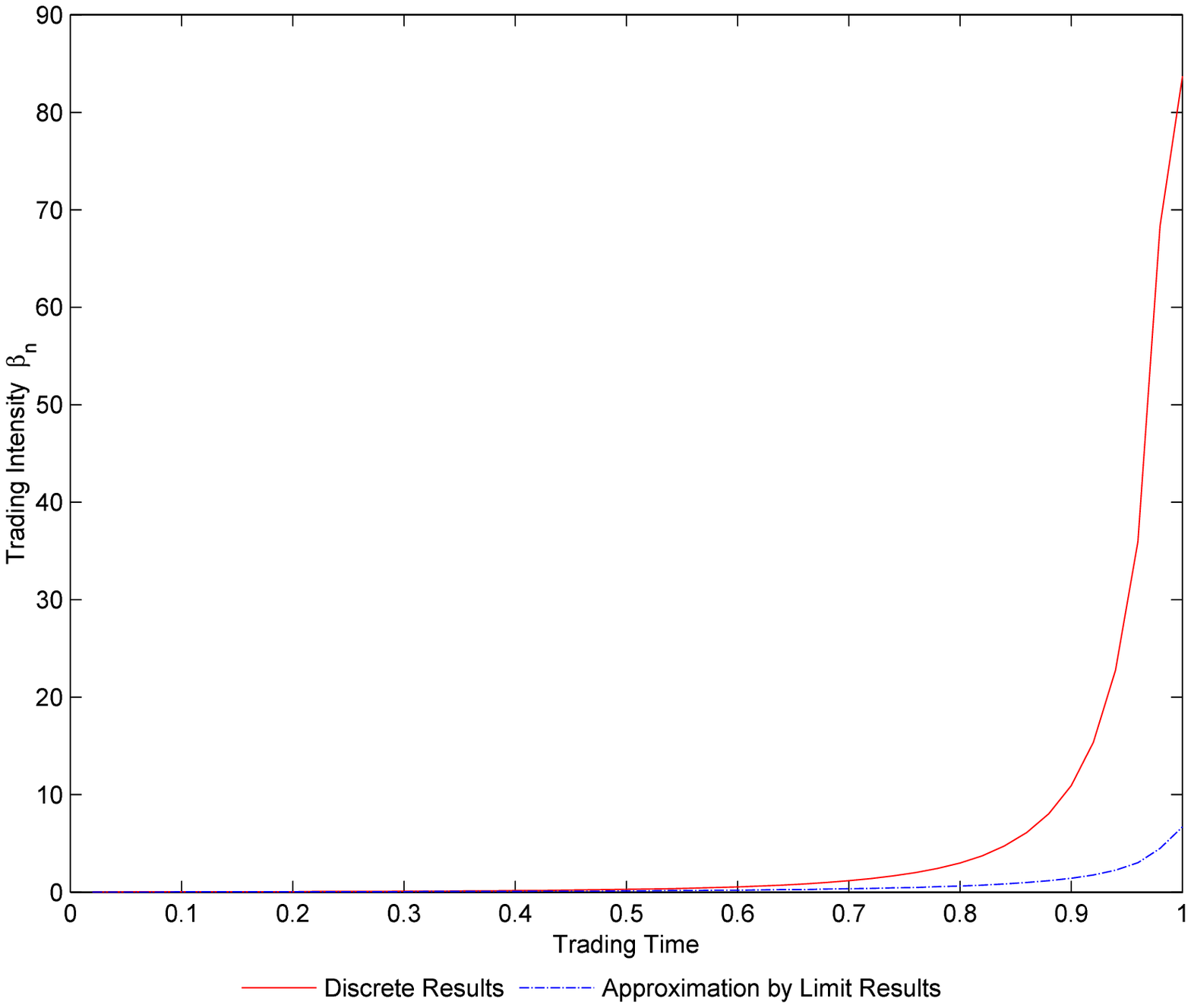}}
} \caption{  (Model 1) Numeric solutions of the intensity trading on
private information $\beta_{n}$ with one unit of initial variance of
information, half unit of noise trader variance across all periods.}
\end{figure}

The specific parameterization that we choose is $\Sigma_{0}=1,
\sigma_{u}=0.5$ and this is also the initialization for the other
two models' simulations. Figures 1, 2, and 3 plot $\Sigma_{n},
\lambda_{n}$ and $\beta_{n}$ respectively with the trading time
horizon fixed between commencement $t=0$ and end of trading $t=1$.
Among them, the subfigures 1(a), 2(a), and 3(a) plot for different
values of period number $N$, while the other subfigures 1(b), 2(b),
and 3(b) show the comparison between discrete model results and
their approximations calculated from the corresponding limit
results.

Figure 1(a) shows the evolution of unrevealed information
$\Sigma_{n}$. For each trading number $N$, $\Sigma_{n}$ declines to
zero very rapidly through time and as the number of periods
increases, the dropping speed increases dramatically. In fact, when
$N=100$, by the 7th period, more than percent 95 information
asymmetry has been eliminated.  Thus, the risk-averse insider
prefers to the trading pattern with trading concentrated at initial
periods to diminish the risk in future profits. Foster and
Viswanathan (1996) also examine this trading pattern numerically,
stemming from the ``rat race" effect among competitive insiders with
high correlated private signals. By contrast, in Model 1, it is the
risk aversion that motivates this pattern, which applies even with a
monopolistic insider setting.

 Figure 2(a) investigates how the market maker responses to the order flow.
The high information asymmetry implies a high adverse selection
early on and the following little unrevealed information implies a
low adverse selection latter. As the number of trading rounds
increasing, the contrast of adverse selections between the early and
the later rounds is more marked since the higher value for the early
rounds and lower for the later is producing a more dramatic decline.

Figure 3(a) shows that $\beta_{n}$ evolves in a manner contrary to
those of $\Sigma_{n}$ and $\lambda_{n}$. For each period number $N$,
$\beta_{n}$ is increasing through all trading periods. As mentioned
earlier, this stems from the decreasing information available and
the decreasing concern for the effect of current trading on future
profits. Moreover, for large $N$, insider's trading on private
information is more intense, with a sharper increase and a higher
terminal value.

At last, figures 1(b), 2(b), 3(b) show that when $N$ is large
sufficient (N=1000, 1000, 50 respectively), the limit results
obtained by asymptotic analysis can give a good characterization to
the actual discrete results.

\begin{center}
 \section{\small{MODEL 2: THE EQUILIBRIUM OF THE RISKY-NEUTRAL INSIDER
 MODEL}}
\end{center}

\subsection {The Discrete Equilibrium}
\setcounter{equation}{0}
 Different with Model 1 in profit-maximization manner, insider in Model 2
 treats
the risky profits and  guaranteed profits equally by maximizing the
sum of their ex ante expectations. The following theorem
characterizes the  equilibrium in discrete case of Model 2.

\begin{theorem}\label{theorem.3}\label{theorem}
In Model 2 with trading period number $N$, a subgame perfect
equilibrium exists. In this equilibrium, there are real numbers
$\beta_{n}, \lambda_{n}, \alpha_{n}$ and $\Sigma_{n}$, such that:
\begin{align}\label{eq.4.1}
&x_{n}=\beta_{n} (v-p_{n-1}), \\
\label{eq.4.2}
&p_{n}-p_{n-1}=\lambda_{n}y_{n},\\
\label{eq.4.3}
&\Sigma_{n}=var (v|y_{1}, y_{2}, \cdots, y_{n}), \\
\label{eq.4.4} &E (\sum^{N}_{k=n}\pi_{k}|p_{1}, p_{2}, \cdots,
p_{n-1}, v) =\alpha_{n-1} (v-p_{n-1}) ^2+\delta_{n-1}.
\end{align}
The above  real numbers  $\beta_{n}, \alpha_{n}$ and $\Sigma_{n}$
can be represented as:
\begin{align}\label{eq.4.5}
&\delta_{n}=a_{n}\sigma_{u}{\Delta{t_{N}}}^{1/2}{\Sigma_{n}}^{1/2}&
(n=0, 1, 2, \cdots, N-1),&\\
\label{eq.4.6}
&\alpha_{n}=b_{n}\sigma_{u}{\Delta{t_{N}}}^{1/2}{\Sigma_{n}}^{-1/2}
 &\quad (n=0, 1, 2, \cdots, N-1),&\\
\label{eq.4.7}
&\beta_{n}=c_{n}\sigma_{u}{\Delta{t_{N}}}^{1/2}{\Sigma_{n-1}}^{-1/2}
& \quad (n=1, 2, \cdots, N).&
\end{align}
in which the sequences $\{a_{n}\}$, $\{b_{n}\}$, $\{c_{n}\}$, with
terminal values $a_{N-1}=0$, $b_{N-1}=\frac{1}{2}$, $c_{N}=1$, are
given recursively: \begin{align}\label{eq.4.8}
&a_{n-1}=a_{n} (\frac{1}{c^2_{n}+1}) ^{1/2}+b_{n} (\frac{1}{c^2_{n}+1}) ^{3/2}c^2_{n},\\
\label{eq.4.9}
&b_{n-1}=b_{n} (\frac{1}{c^2_{n}+1}) ^{3/2}+\frac{c_{n}}{c^2_{n}+1},\\
\label{eq.4.10} &a_{n}+b_{n}=\frac{1-c^2_{n}}{c_{n} (1+c^2_{n})
^{1/2}},
\end{align}
where $c_{n}>0, n=1, 2, \cdots, N$.
\end{theorem}

Compared with Theorem 1 in Model 1, the only difference literally is
insider's strategy formulation (\ref{eq.4.10}). Inspecting it shows
the insider's  trading intensity coefficient $c_{n}$ is a decreasing
function of $a_{n}+b_{n}$. This means that if insider has more
ability to earn the ex ante expectation of future profits that
cumulated from the next period to the end, then she will trade less
aggressively at the current period to keep more information
advantage for the next period.

 Specifically, as in Model 1, we investigate the two periods case for Model 2 (with endogenous parameters added an upper index (2)) and
compare our results to those of Kyle (1985) (with endogenous
parameters added an upper index (0)). By (\ref{eq.3.210}), the ex
ante expectation of profits in Kyle (1985) satisfies
\begin{align*}
E[\sum^{2}_{i=1}\pi^{ (0) }_{i}]=\alpha^{ (0)
}_{0}\Sigma_{0}+\delta^{ (0) }_{0}\approx0.8776
\Sigma^{1/2}_{0}\sigma_{u}{{\Delta}t_{N}}^{1/2}.
\end{align*}
In Model 2, the endogenous parameters result in a larger ex ante
expectation of profits, i.e.,
\begin{align*}
E[\sum^{2}_{i=1}\pi^{ (2) }_{i}]=\alpha^{ (2)
}_{0}\Sigma_{0}+\delta^{ (2) }_{0}>E[\sum^{2}_{i=1}\pi^{ (0) }_{i}]
\end{align*}with
\begin{align*}
\alpha^{ (2) }_{0}\approx 0.7385
\Sigma^{-1/2}_{0}\sigma_{u}{{\Delta}t_{N}}^{1/2}, \qquad \delta^{
(2) }_{0}\approx 0.1416
\Sigma^{1/2}_{0}\sigma_{u}{{\Delta}t_{N}}^{1/2}.
\end{align*}
Moreover, we have,
\begin{align*}
\beta^{ (2) }_{1}>\beta^{ (0) }_{1}, \ \beta^{ (2) }_{2}>\beta^{ (0)
}_{2}, \ \Sigma^{ (2) }_{1}<\Sigma^{ (0) }_{1}, \ \Sigma^{ (2)
}_{2}<\Sigma^{ (0) }_{2}, \ \lambda^{ (2) }_{1}>\lambda^{ (0) }_{1},
\ \lambda^{ (2) }_{2}<\lambda^{ (0) }_{2}.
\end{align*}
Thus, the insider trades more aggressively, reveals more
information, and induces a higher adverse selection in the initial
period and a lower  adverse selection in the last period in Model 2
than in Kyle (1985) model. These results can generalize to the
N-period case which we will show numerically.

\subsection {The Limit Behavior When $N\rightarrow\infty$}

Similarly, we need some preliminary results for the limit results,
given by following.
\begin{proposition}\label{pro.4}
The sequence $\{c_{n}\}$ in Theorem 3 can be achieved as follows.
 Given $c_{n}$ , $c_{n-1}$ is determined  by
 the unique root lies in $ (0, 1) $ of equation
\begin{align}\label{eq.4.11}
 (1-c^2_{n-1}) c_{n} (1+c^2_{n}) =c_{n-1} (1+c^2_{n-1}) ^{1/2}
\end{align}
with terminal values $c_{N}=1$. Moreover, the following monotonicity
holds:
\begin{align}\label{eq.4.12}
c_{N}>c_{N-1}>\cdots, c_{n}>c_{n-1}>c_{2}>c_{1}.\end{align}
\end{proposition}

With the Proposition \ref{pro.3}, the limit of discrete equilibrium
of Model 2 follows. Unlike in Model 1, we focus on the first class
of limits since the limit equilibrium exists.
\begin{theorem}\label{theorem.4}
When $N\rightarrow\infty$, the sequences $\{a_{n}\}, \{b_{n}\}$ and
$\{c_{n}\}$ in Theorem \ref{theorem.3} have limits:
\begin{align}\label{eq.4.13}
&\lim\limits_{N\rightarrow\infty}c_{[Nt]}=0, \\
\label{eq.4.14}
&\lim\limits_{N\rightarrow\infty}b_{[Nt]}=\infty, \\
\label{eq.4.15} &\lim\limits_{N\rightarrow\infty}a_{[Nt]}=\infty,
\end{align} for any $t\in (0, 1) $.
Moreover,
\begin{align}\label{eq.4.16}
&\lim\limits_{N\rightarrow\infty}\frac{c_{[Nt]}}{{\Delta{t}}^{1/2}_{N}}=\frac{1}{ (1-t) ^{1/2}}, \\
\label{eq.4.17}
&\lim\limits_{N\rightarrow\infty}b_{[Nt]}{\Delta{t}}^{1/2}_{N}=\frac{1}{2} (1-t) ^{1/2},\\
\label{eq.4.18}
&\lim\limits_{N\rightarrow\infty}a_{[Nt]}{\Delta{t}}^{1/2}_{N}=\frac{1}{2}
 (1-t) ^{1/2}.
\end{align}
Insider dissipates her private information gradually, that is,
\begin{align}\label{eq.4.19}&\lim\limits_{N\rightarrow\infty}\Sigma_{[Nt]}=
 (1-t) \Sigma_{0},\\
\label{eq.4.20}&\lim\limits_{N\rightarrow\infty}\lambda_{[Nt]}=\frac{\Sigma^{1/2}_{0}}{\sigma_{u}},\\
\label{eq.4.21}&\lim\limits_{N\rightarrow\infty}\frac{\beta_{[Nt]}}{\Delta{t_{N}}}=\frac{\sigma_{u}}{
(1-t) \Sigma^{1/2}_{0}}.\end{align}
\end{theorem}

From Theorem 4, the discrete equilibrium results of Model 2 converge
to the continuous  equilibrium results of  Kyle (1985). Therefore,
 as trading happens more and more frequently, the difference between equilibriums in discrete case when insider takes the
pricing rule as can and cannot be influenced disappears. This fact
might stem form the continuity of liquidity parameter when trading
happens continuously in the sense the liquidity parameter that will
arise can be deduced accurately from the former level, regardless of
whether or not insider thinks she can influence it.

\subsection {Numerical Results}

\begin{figure}
\centering \mbox{ \subfigure[Unrevealed information
$\Sigma_{n}$.]{\includegraphics[width=.32\textwidth]{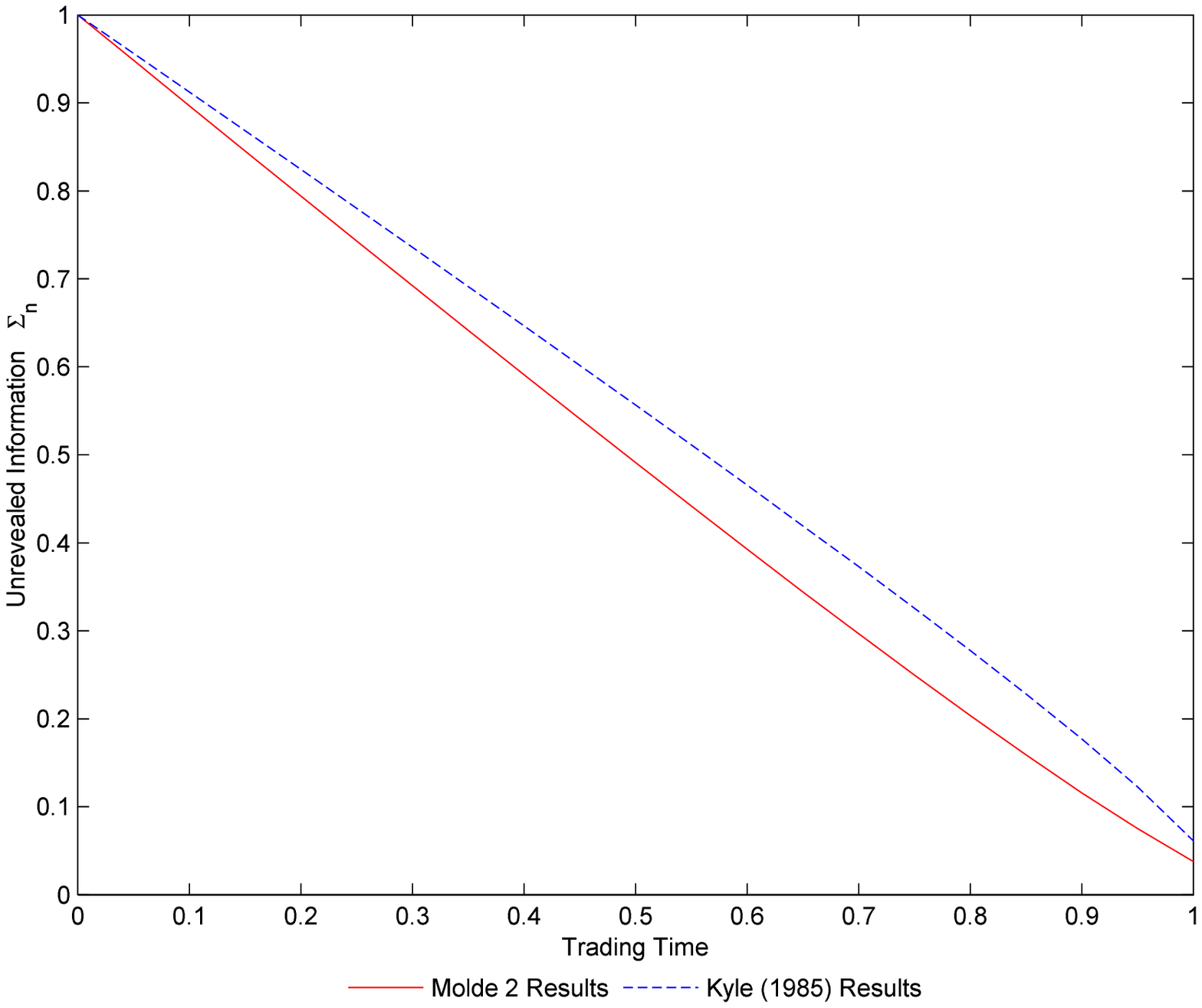}}\quad
 \subfigure[Liquidity parameter $\lambda_{n}$.]{\includegraphics[width=.32\textwidth]{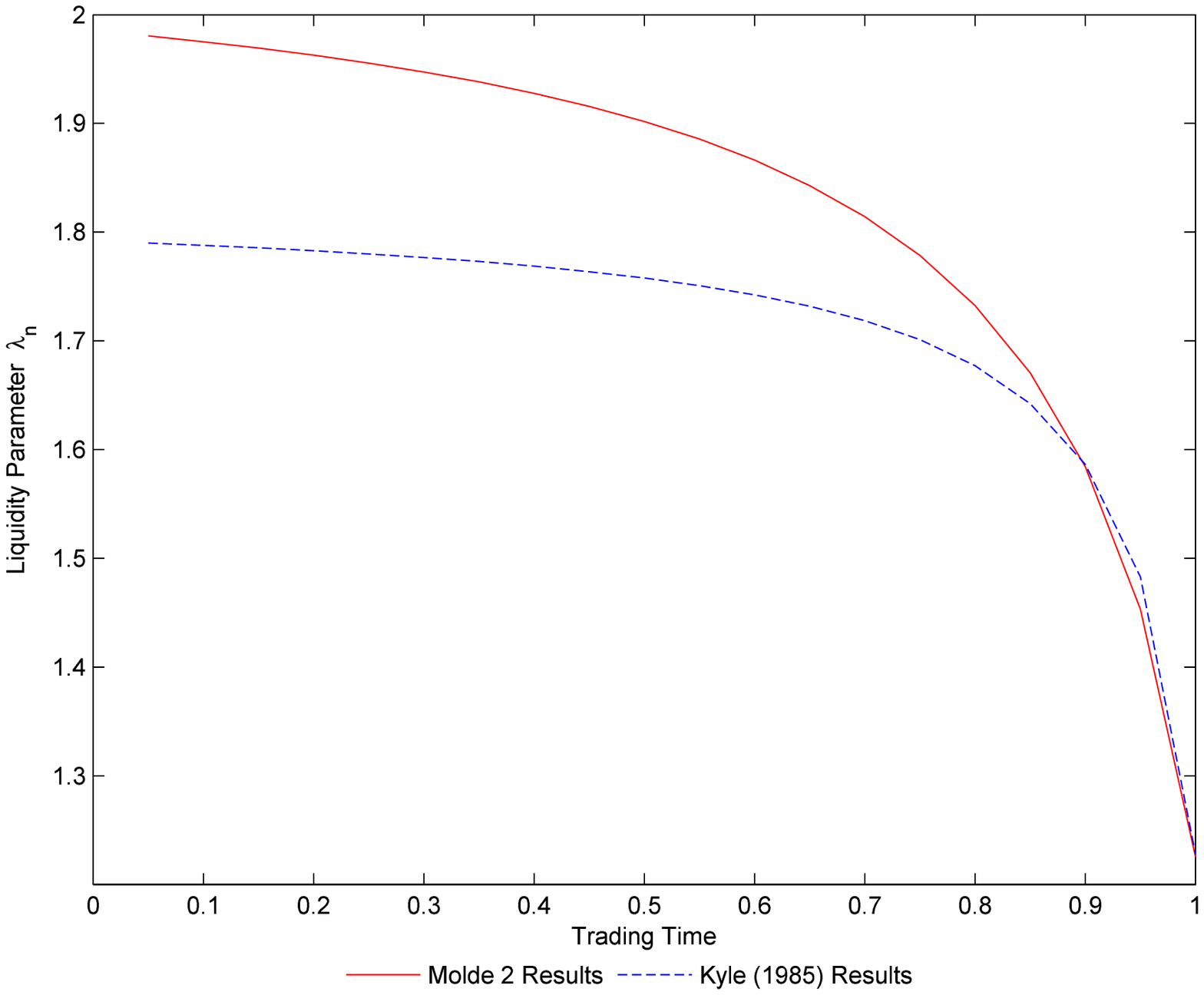}}
 \quad
 \subfigure[Trading intensity
 $\beta_{n}$.]{\includegraphics[width=.32\textwidth]{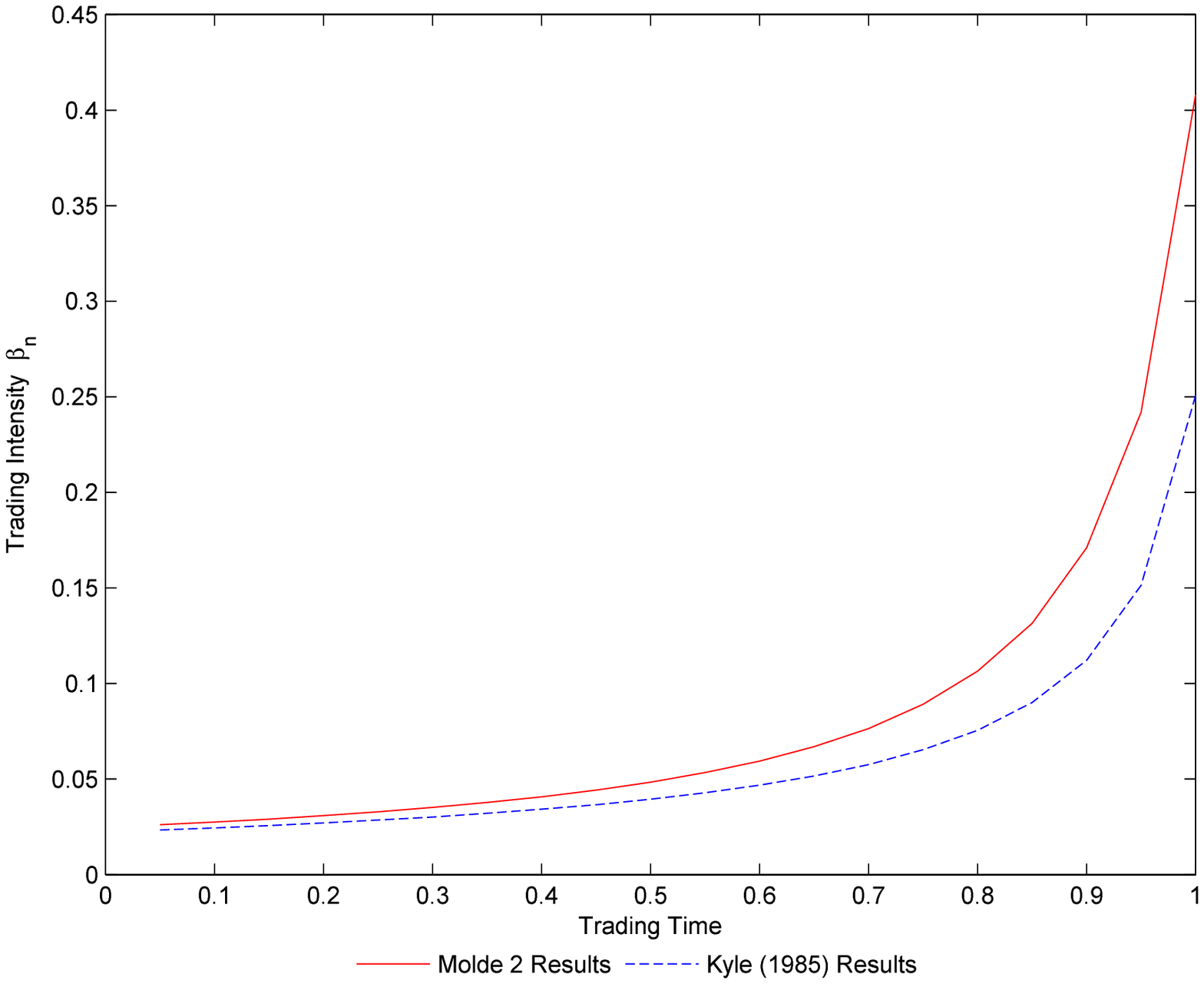}}}
 \caption{Equilibriums in Model 2 and Kyle (1985) model when
N=20.}
\end{figure}
\begin{figure}
\centering \mbox{ \subfigure[Unrevealed information $\Sigma_{n}$
with changing period numbers $N=5, N=20,
N=100$.]{\includegraphics[width=.47\textwidth]{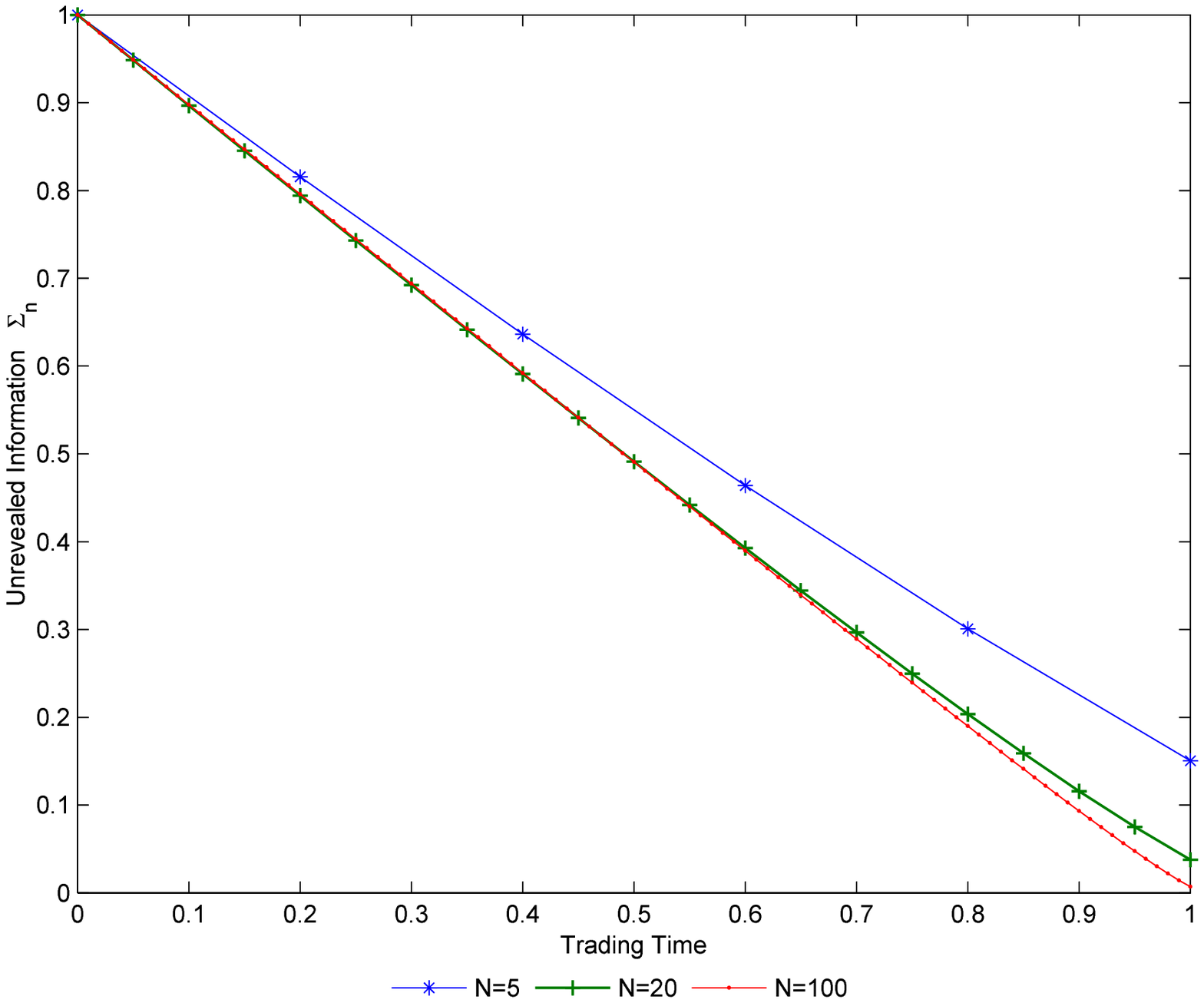}
 }\qquad
 \subfigure[ The approximation by limit results and the actual discrete results when N=100.]{\includegraphics[width=.47\textwidth]{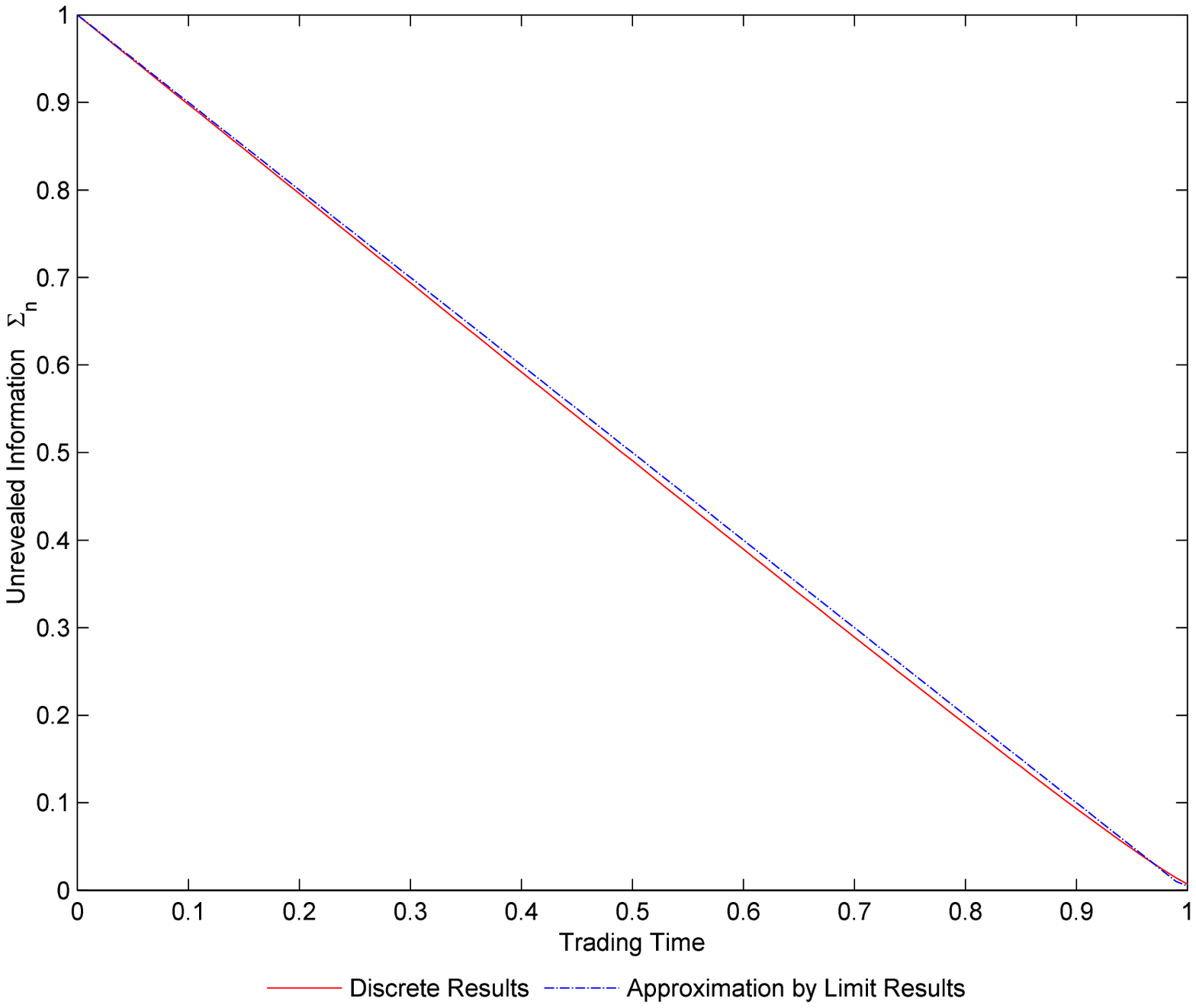}}
} \caption{ (Model 2) Numeric solutions of the unrevealed
information $\Sigma_{n}$ with one unit of initial variance of
information, half unit of noise trader variance across all periods.}
\end{figure}
\begin{figure}
\centering \mbox{ \subfigure[Liquidity parameter $\lambda_{n}$ with
changing period numbers $N=5, N=20,
N=100$.]{\includegraphics[width=.47\textwidth]{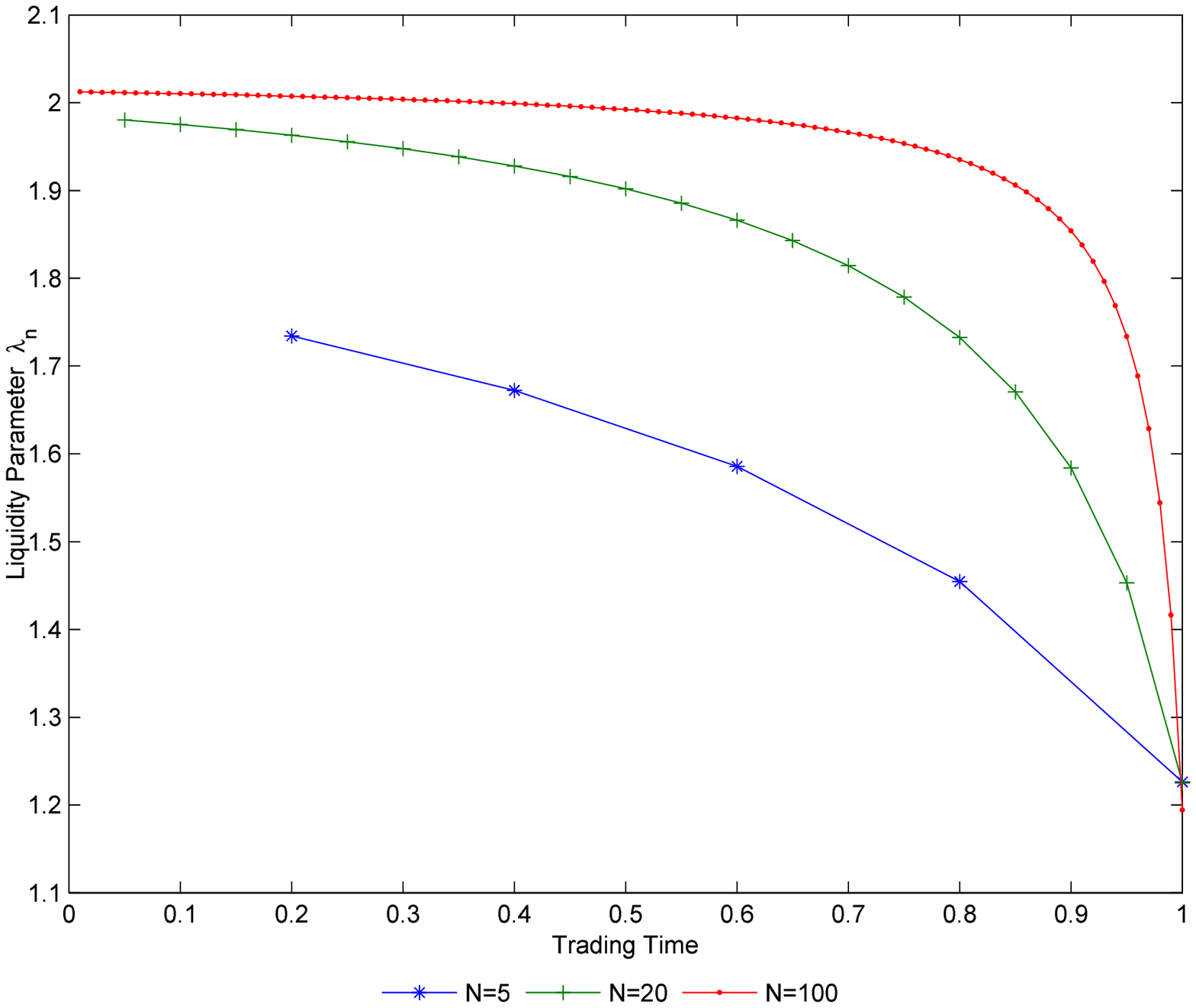}
 }\qquad
 \subfigure[ The approximation by limit results and the actual discrete results when N=100. ]{\includegraphics[width=.47\textwidth]{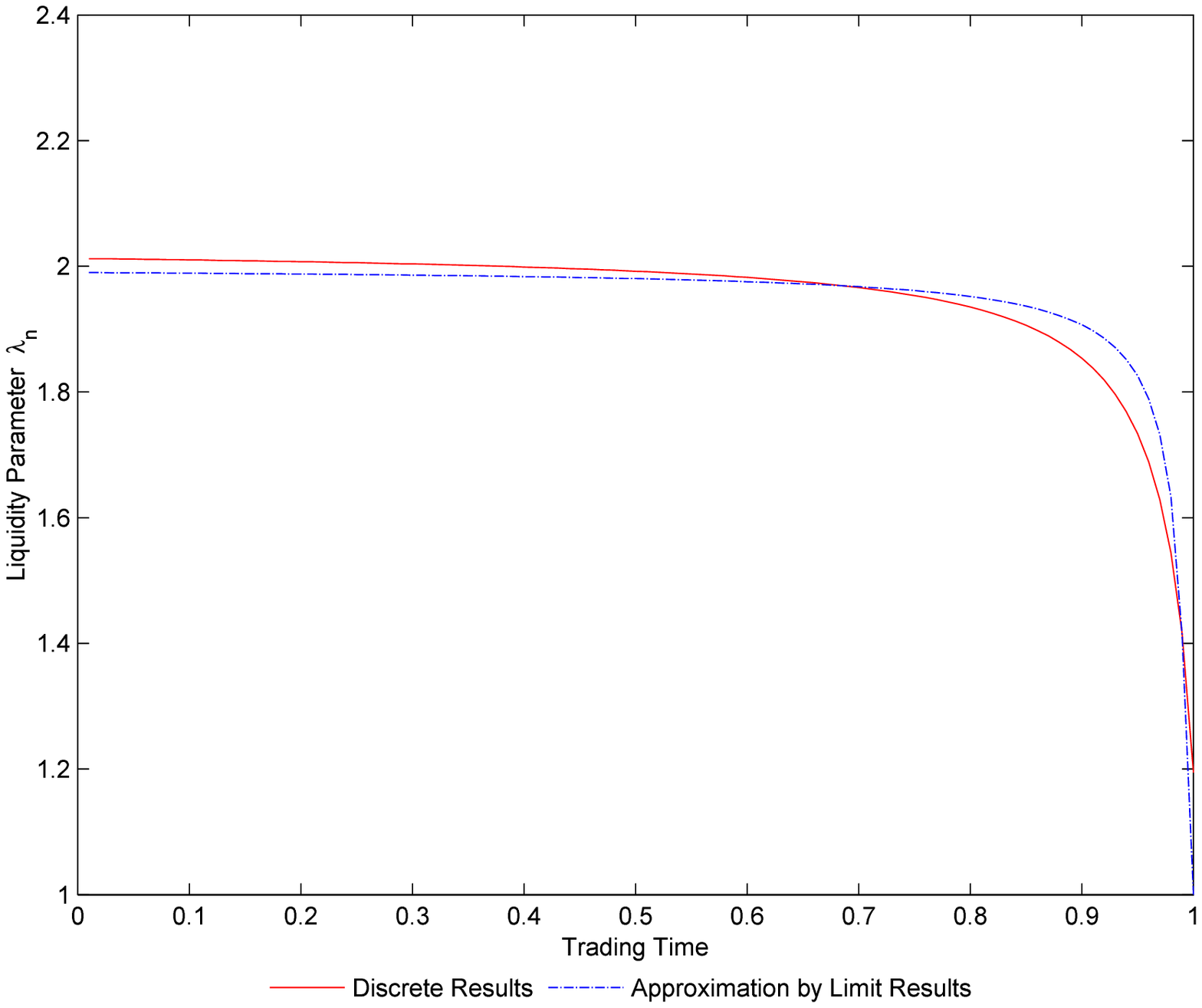}}
} \caption{ (Model 2) numeric solutions of liquidity parameter
$\lambda_{n}$ with one unit of initial variance of information, half
unit of noise trader variance across all periods.}
\end{figure}
\quad
\begin{figure}
\centering \mbox{ \subfigure[Intensity trading on private
information $\beta_{n}$ with changing period numbers $N=5, N=20,
N=100$.]{\includegraphics[width=.47\textwidth]{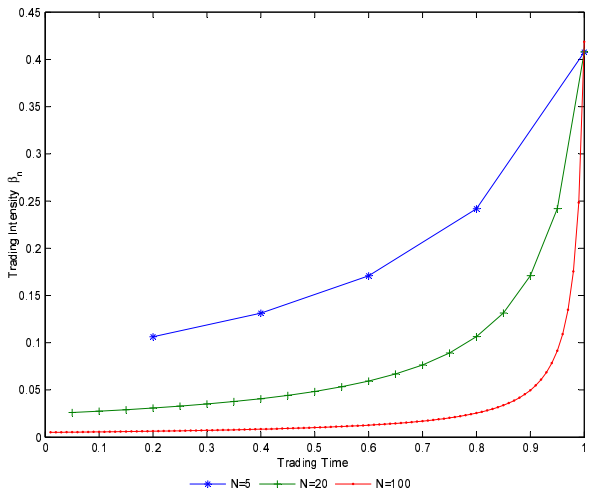}
 }\qquad
 \subfigure[  The approximation by limit results and the actual discrete results when N=100. ]{\includegraphics[width=.47\textwidth]{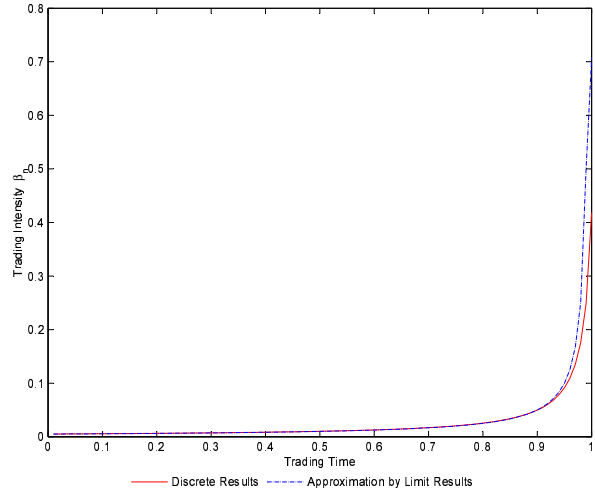}}
} \caption{  (Model 2) numeric solutions of intensity trading on
private information $\beta_{n}$ with one unit of initial variance of
information, half unit of noise trader variance across all periods.}
\end{figure}

Since the Model 2 has the same limit equilibrium as Kyle (1985), it
is necessary to consider the discrete case depicted numerically.
Figure 4 shows that compared to  Kyle (1985), the insider in Model 2
trades more aggressively on private information (Figure 4(c)) and
reveals more information by any time $t>0$ (Figure 4(a)). The more
aggressive trading in Model 2 induces a higher adverse selection for
a long time, and then results in a slightly lower adverse in the
last few periods due to the less scale of unrevealed information
(Figure 4(b). Note also that $\lambda_{n}\sigma^2_{u}$ represents
the expected liquidity cost or informed profits in the nth period.
Thus, subfigure 4(b)  implies that the insider in Model 2 can obtain
a higher ex ante expectation of total profits across all periods
compared to that in Kyle (1985).

Figure 5(a) shows that the risk-neutral insider exploits  her
private information slowly, at an almost constant speed even when
there are only 5 trading opportunities. Additionally, a large
trading opportunity implies more information being revealed
publicly.

 Figure 6(a) indicates that the adverse selection curve is
concave, in contrast to the convex adverse selection curve in Model
1. Moreover, although the adverse selection is decreasing in both
models, the most decline here happens approaching the end of
trading, while in Model 1 it happens in initial periods. Another
difference is that here the larger trading opportunity always
results in a higher adverse selection before the end of trading,
whereas it usually implies a lower adverse selection in Model 1.

Figure 7(a) shows that, like  Model 1,  for each period number $N$,
$\beta_{n}$ is increasing with time going on, and in contrast to
Model 1, with large trading opportunity, insider here would like to
trade more softly on private information than the case with small
trading opportunity for all periods except the last one.

 Finally,
Figures 5(b), 6(b), 7(b) show that when $N$ is 100, the limit
results obtained by asymptotic analysis can characterize the actual
discrete results well.

\begin{center}  \section{\small{MODEL 3: THE EQUILIBRIUM OF THE RISK-SEEKING INSIDER
 MODEL}}
\end{center}

\subsection {The Discrete Equilibrium}
\setcounter{equation}{0}

 The risk-seeking insider in Model 3
behaves oppositely to the risk-averse insider in Model 1. The
following theorem characterizes the equilibrium in discrete case of
Model 3, with insider's strategies distinct from those in the former
two models.

\begin{theorem}\label{theorem.5}
In Model 3 with trading period number $N$, a subgame perfect linear
equilibrium exists. In this equilibrium, there are real numbers
$\beta_{n}, \lambda_{n}, \alpha_{n}$ and $\Sigma_{n}$, such that:
\begin{align}\label{eq.5.1}
&x_{n}=\beta_{n} (v-p_{n-1}) \\
\label{eq.5.2}
&p_{n}-p_{n-1}=\lambda_{n}y_{n}\\
\label{eq.5.3}
&\Sigma_{n}=var (v|y_{1}, y_{2}, \cdots, y_{n}) \\
\label{eq.5.4} &E (\sum^{N}_{k=n}\pi_{k}|p_{1}, p_{2}, \cdots,
p_{n-1}, v) =\alpha_{n-1} (v-p_{n-1}) ^2+\delta_{n-1}
\end{align}
The above  real numbers  $\beta_{n}, \alpha_{n}$ and $\Sigma_{n}$
can be represented as:
\begin{align}\label{eq.5.5}
&\delta_{n}=a_{n}\sigma_{u}{\Delta{t_{N}}}^{1/2}{\Sigma_{n}}^{1/2}&
(n=0, 1, 2, \cdots, N-1),&\\
\label{eq.5.6}
&\alpha_{n}=b_{n}\sigma_{u}{\Delta{t_{N}}}^{1/2}{\Sigma_{n}}^{-1/2}
 &\quad (n=0, 1, 2, \cdots, N-1),&\\
\label{eq.5.7}
&\beta_{n}=c_{n}\sigma_{u}{\Delta{t_{N}}}^{1/2}{\Sigma_{n-1}}^{-1/2}
& \quad (n=1, 2, \cdots, N).&
\end{align}
in which the sequences $\{a_{n}\}$, $\{b_{n}\}$, $\{c_{n}\}$, with
terminal values $a_{N-1}=0$, $b_{N-1}=\frac{1}{2}$, $c_{N}=1$, are
given recursively: \begin{align}\label{eq.5.8}
&a_{n-1}=a_{n} (\frac{1}{c^2_{n}+1}) ^{1/2}+b_{n} (\frac{1}{c^2_{n}+1}) ^{3/2}c^2_{n}\\
\label{eq.5.9}
&b_{n-1}=b_{n} (\frac{1}{c^2_{n}+1}) ^{3/2}+\frac{c_{n}}{c^2_{n}+1}\\
\label{eq.5.10} &-3b_{n}c_{n}+ (c^2_{n}+1) ^{1/2} (1-c^2_{n}) =0
\end{align} where $c_{n}>0, n=1, 2, \cdots, N$.
\end{theorem}

 From (\ref{eq.5.10}), we find that in the nth period, the insider's  trading intensity coefficient $c_{n}$
 is decreasing with the marginal risky profits estimated in the n+1th period $b_{n}$.
 This reflects the fact, in contrast to that in Model 1,
 the risk-seeking insider has an incentive to trade less currently to keep more
 information for the next period, provided her with more ability to
 earn risky profits in  the next period.

Like the former 2 models, we can solve the equilibrium easily when
$N=2$. In Model 3, we add an upper index (3) to the endogenous
parameters. Calculations show:
\begin{align*}
\alpha^{ (3) }_{0}\approx 0.7587
{\Sigma_{0}}^{-1/2}\sigma_{u}{\Delta t_{N}}^{1/2}, \qquad \delta^{
(0) }_{0}\approx 0.0988 {\Sigma_{0}}^{1/2}\sigma_{u}{\Delta
t_{N}}^{1/2}.
\end{align*}
Therefore, at the beginning of trading, the risk-seeking insider in
Model 3 obtains larger risky profits at the cost of  smaller
guaranteed profits relative to the Model 1, 2 and Kyle (1985) model.

\subsection {The Limit Behavior When $N\rightarrow\infty$}

Similarly, in Model 3, we have:

 \begin{proposition}\label{pro.5}
The sequence $\{c_{n}\}$ in Theorem 5 can be achieved as follows.
 Given $c_{n}$ , $c_{n-1}$ is determined  by
 the unique root lies in $ (0, 1) $ of equation
\begin{align}\label{eq.5.11}
\frac{1}{c_{n-1}} (c^2_{n-1}+1) ^{1/2} (1-c^2_{n-1})
=\frac{1}{c^2_{n}+1} (\frac{1}{c_{n}}+2c_{n})
\end{align}
with terminal values $c_{N}=1$. Moreover, the following monotonicity
holds:
\begin{align}\label{eq.5.12}
c_{N}>c_{N-1}>\cdots, c_{n}>c_{n-1}>c_{2}>c_{1},\end{align} and
sequence $\{b_{n}\}$ has an expression of sequence $\{c_{n}\}$
\begin{align}\label{eq.5.13}
b_{n}=\frac{ (c^2_{n}+1) ^{1/2}}{3c_{n}} (1-c^2_{n}).
\end{align}
\end{proposition}

\begin{theorem}\label{theorem.6}
When $N\rightarrow\infty$, the sequences $\{a_{n}\}, \{b_{n}\}, $
and $\{c_{n}\}$ in Theorem 5 have limits:
\begin{align}\label{eq.5.14}
&\lim\limits_{N\rightarrow\infty}c_{[Nt]}=0, \\
\label{eq.5.15}
&\lim\limits_{N\rightarrow\infty}b_{[Nt]}=\infty, \\
\label{eq.5.16} &\lim\limits_{N\rightarrow\infty}a_{[Nt]}=\infty,
\end{align} for any $t\in (0, 1) $.
Moreover,
\begin{align}\label{eq.5.17}
&\lim\limits_{N\rightarrow\infty}\frac{c_{[Nt]}}{{\Delta{t}}^{1/2}_{N}}=\frac{\sqrt{3}}{3}\frac{1}{ (1-t) ^{1/2}}, \\
\label{eq.5.18}
&\lim\limits_{N\rightarrow\infty}b_{[Nt]}{\Delta{t}}^{1/2}_{N}=\frac{\sqrt{3}}{3} (1-t) ^{1/2},\\
\label{eq.5.19}
&\lim\limits_{N\rightarrow\infty}a_{[Nt]}{\Delta{t}}^{1/2}_{N}=\frac{\sqrt{3}}{6}
 (1-t) ^{1/2}.
\end{align}
The insider exploits her private information slowly, that is,
\begin{align}\label{eq.5.20}&\lim\limits_{N\rightarrow\infty}\Sigma_{[Nt]}= (1-t)
^{1/3}\Sigma_{0},\\
\label{eq.5.21}
&\lim\limits_{N\rightarrow\infty}\lambda_{[Nt]}=\frac{\sqrt{3}\Sigma_{0}^{1/2}}{3
(1-t) ^{1/3}\sigma_{u}},\\
\label{eq.5.22}&\lim\limits_{N\rightarrow\infty}\frac{\beta_{[Nt]}}{\Delta{t_{N}}}=
\frac{\sqrt{3}\sigma_{u}}{3 (1-t)
^{2/3}\Sigma^{1/2}_{0}}.\end{align}
\end{theorem}
As shown by Theorem 6, the information revealing speed, liquidity
parameter and trading intensity in limit are all increasing with
calender time t. Compared to the corresponding limit results in Kyle
(1985) (or Model 2), the risk-seeking insider in model 3 always
trades less aggressively, always acquires larger risky profits( and
smaller guaranteed profits at the beginning of trading), and  by any
time $t>0$ reveals less information. When $t<1-\frac{\sqrt{3}}{3}$,
the adverse selection is lower than that in Kyle (1985) due to the
low trading intensity while when $t>1-\frac{\sqrt{3}}{3}$, the
adverse selection is higher than that in Kyle (1985) due to the
large scale of unrevealed information. Moreover, $ (\ref{eq.5.18}) $
and $ (\ref{eq.5.19}) $ imply that insider's ability of earning
risky profits is about twice that of earning guaranteed profits,
whereas it is ``half" in Model 1 and ``equal to" in Model 2.

\subsection {Numerical Results}

\begin{figure}
\centering \mbox{ \subfigure[Unrevealed information $\Sigma_{n}$
with changing period numbers $N=5, N=20,
N=100$.]{\includegraphics[width=.47\textwidth]{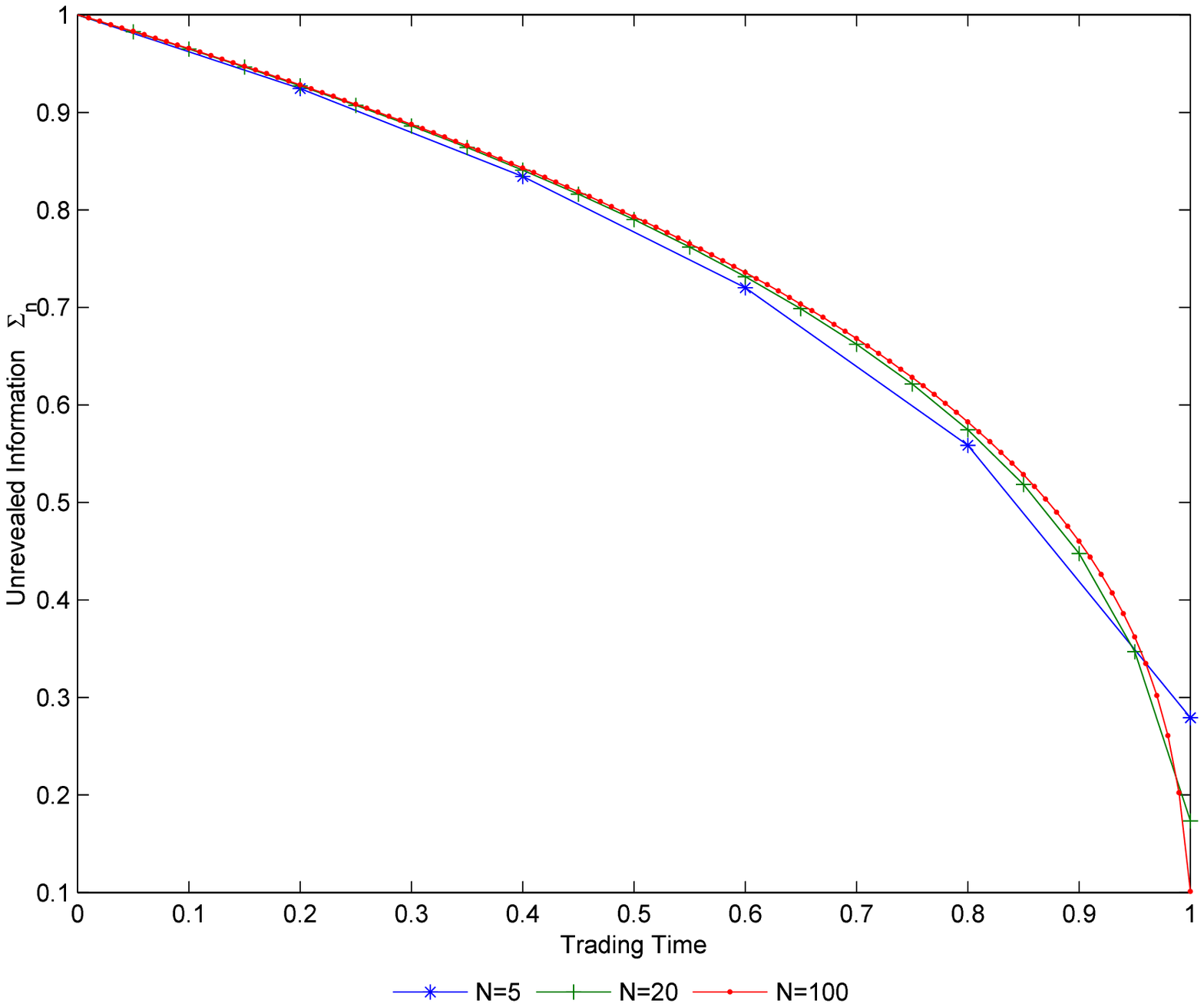}
 }\qquad
 \subfigure[ The approximation by limit results and the actual discrete results when N=20.]{\includegraphics[width=.47\textwidth]{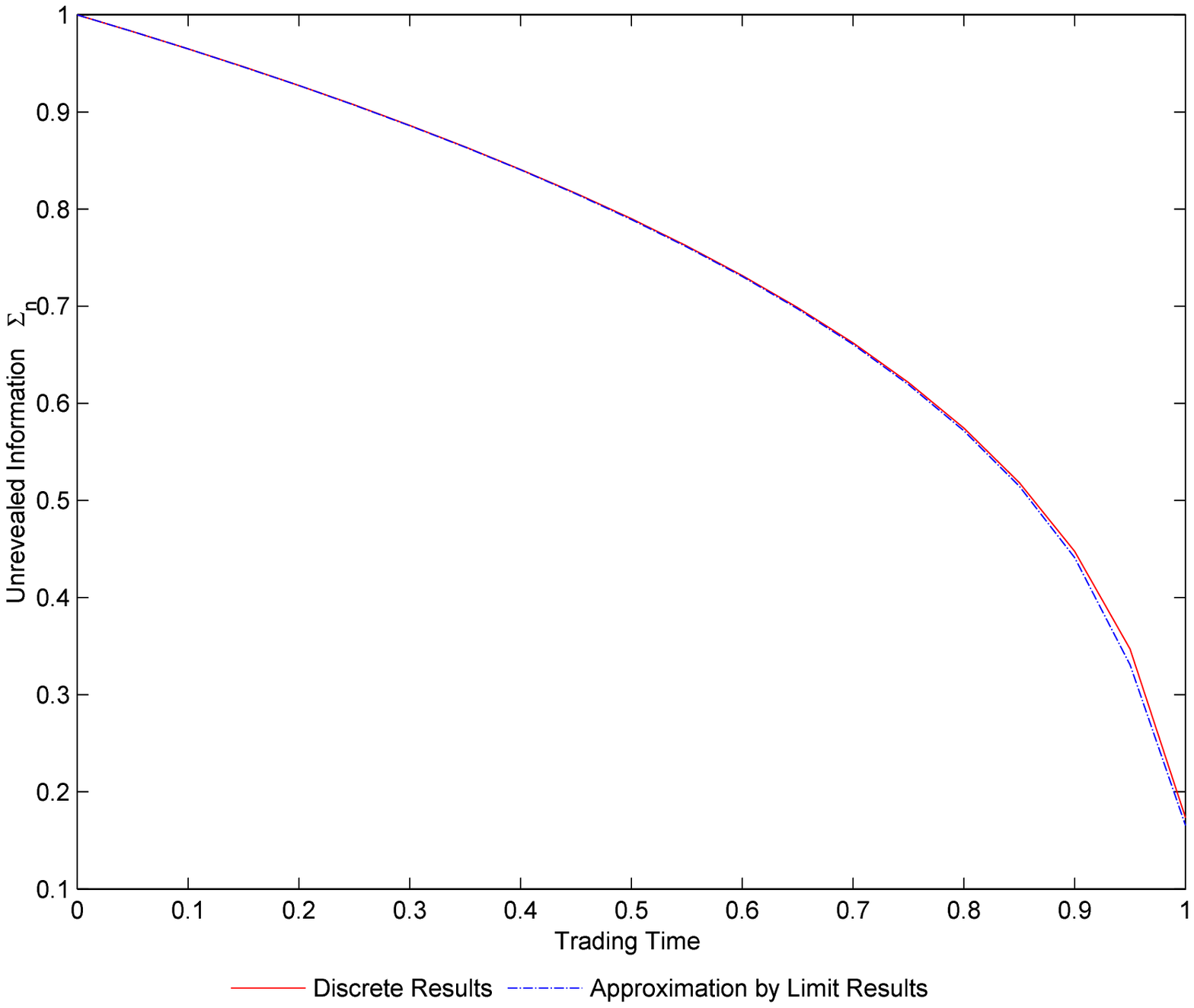}}
} \caption{ (Model 3) Numeric solutions of the unrevealed
information $\Sigma_{n}$ with one unit of initial variance of
information, half unit of noise trader variance  across all
periods.}
\end{figure}
\begin{figure}
\centering \mbox{ \subfigure[Liquidity parameter $\lambda_{n}$ with
changing period numbers $N=5, N=20,
N=100$.]{\includegraphics[width=.47\textwidth]{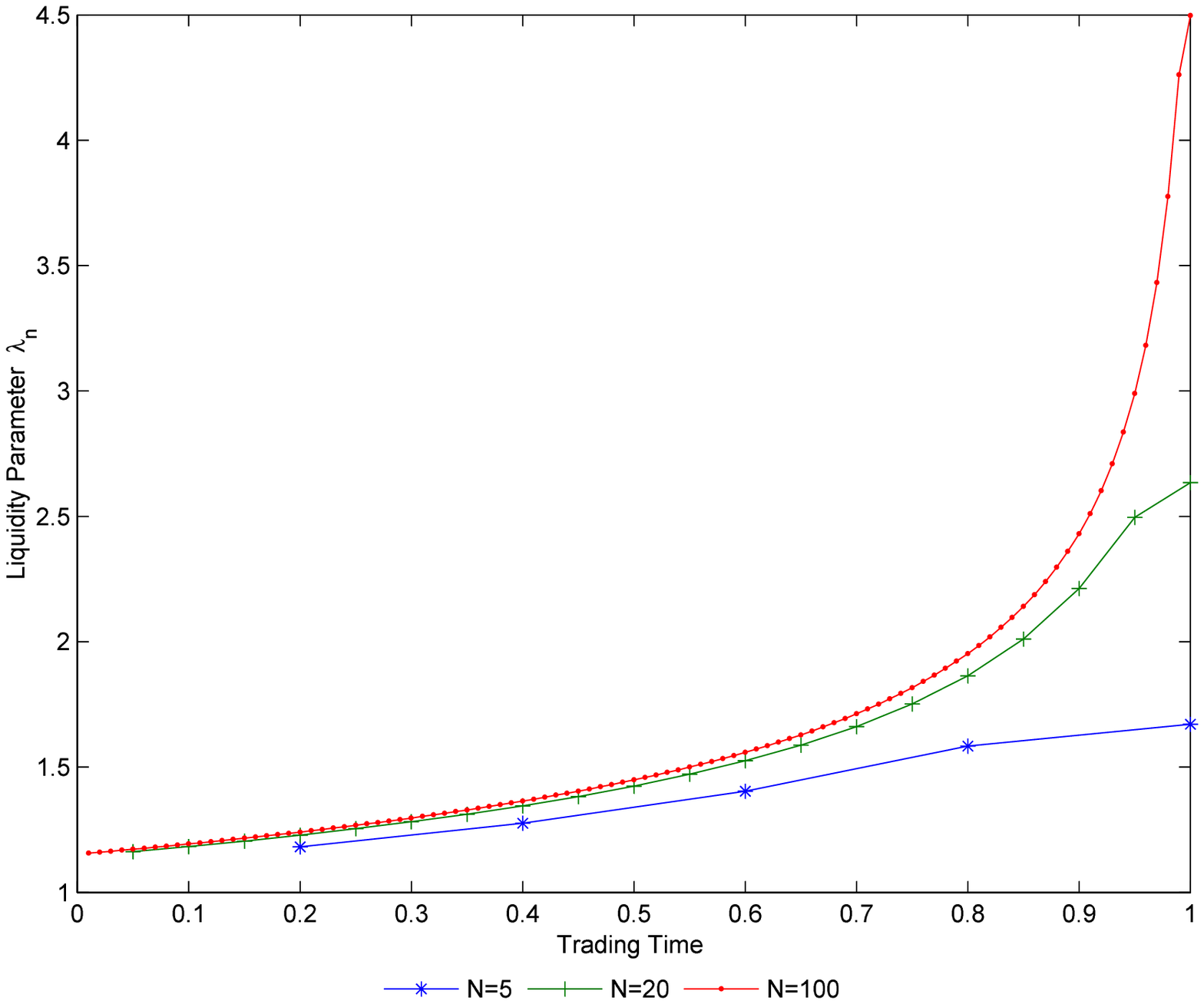}
 }\qquad
 \subfigure[ The approximation by limit results and the actual discrete results when N=20. ]{\includegraphics[width=.47\textwidth]{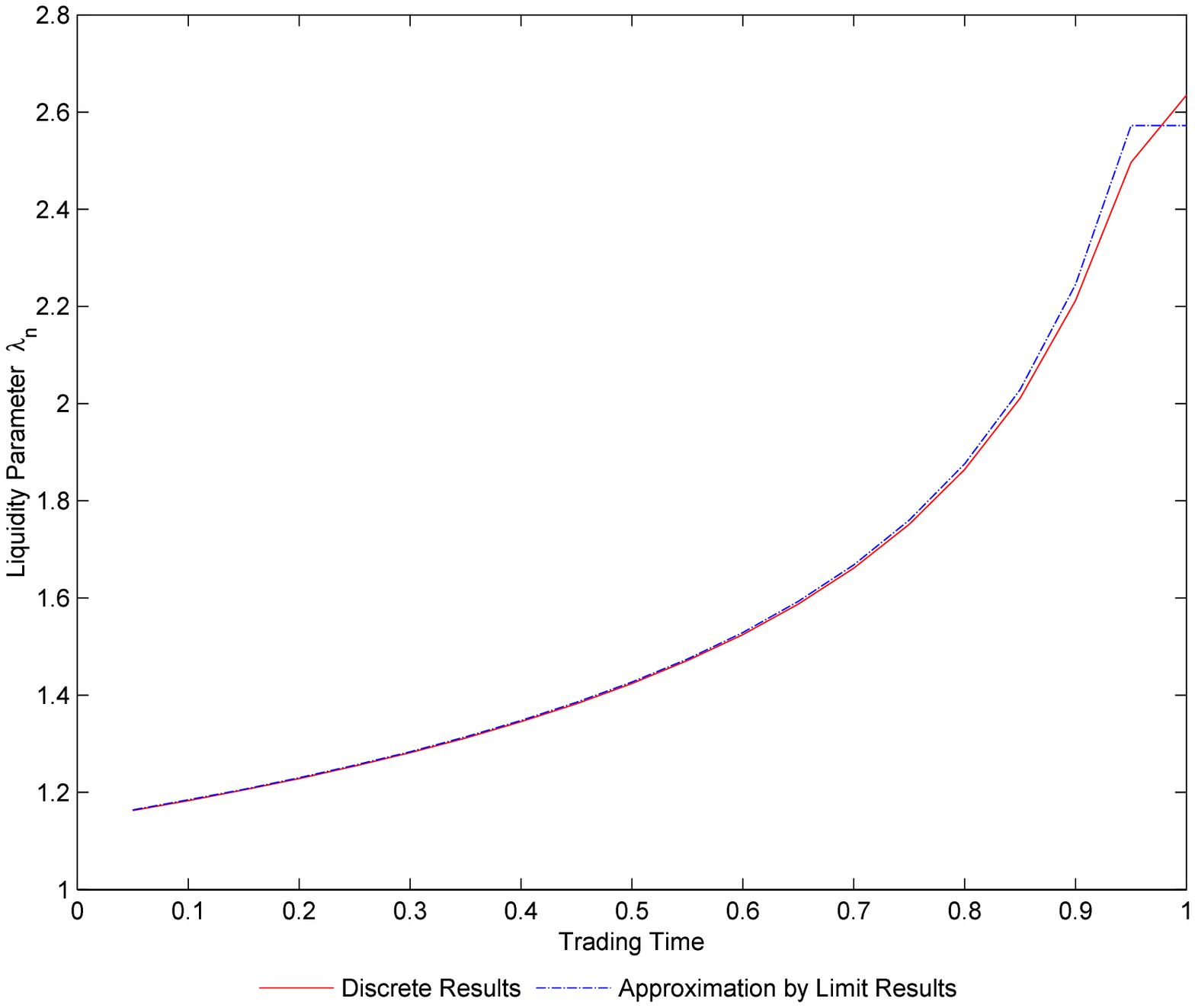}}
} \caption{ (Model 3) numeric solutions of liquidity parameter
$\lambda_{n}$ with one unit of initial variance of information, half
unit of noise trader variance across all periods.}
\end{figure}
\begin{figure}
\centering \mbox{ \subfigure[Intensity trading on private
information $\beta_{n}$ with changing period numbers $N=5, N=20,
N=100$.]{\includegraphics[width=.47\textwidth]{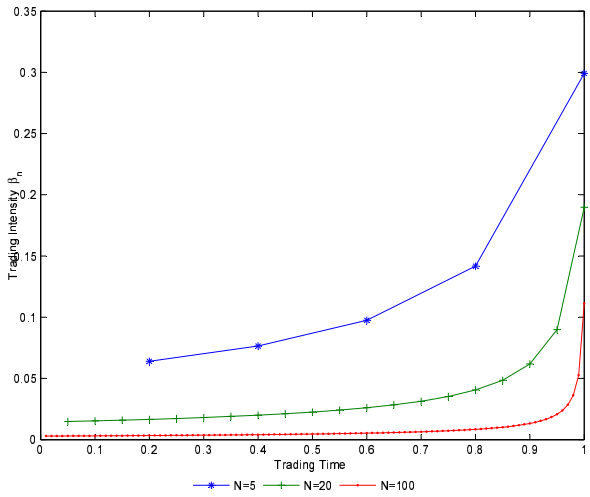}
 }\qquad
 \subfigure[  The approximation by limit results and the actual discrete results when N=20. ]{\includegraphics[width=.47\textwidth]{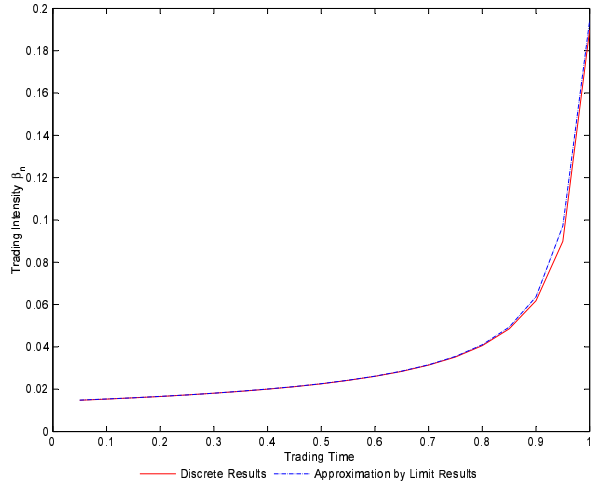}}
} \caption{ (Model 3) numeric solutions of intensity trading on
private information $\beta_{n}$ with one unit of initial variance of
information, half unit of noise trader variance across all periods.}
\end{figure}
Figure 8(a) depicts that the information released to price is of the
least scale among our three models by any time $t>0$. At the initial
periods, the decline speed is low and only when time approaching the
end, the speed is accelerated to exploit the large information
remained unrevealed. Actually, the risk-seeking insider would like
to maintain her information advantage to the end to acquire high
future profits with high risk. These results are actually implying
the ``waiting game" effect once investigated by Foster and
Viswanathan (1996) in which insiders make relatively small trades
for the current periods and large for future, with the hope that the
others would push the price to a wrong direction in the future. By
contrast, the motivation here is the risk-seeking behavior of a
monopolistic insider  while it is the negative related signals
endowed to competitive insiders there.

Figure 9(a) shows for each period number $N$, the adverse selection
is increasing with trades, and this is in sharp contrast to the
other models with decreasing adverse selection. This increase
pattern is consistent with Figure 7(a)'s implication that most
amount of the private information is dissipated at the latter
trading periods. Moreover, for a large $N$, the market maker's
sensitivity to order flow is higher and the increasing is more
pronounced.

Figure 10(a) shows for every
 $N$, insider's trading intensity is increasing through all trading periods.
Specifically, $\beta_{n}$
 is lower as more trading periods are added, different with that in Model
1 and same with Model 2.

At last, Comments about Figures 8(b), 9(b), 10(b) refers to the
other two models.

\begin{center}
\section{Conclusion}
\end{center}

 We improve the Kyle (1985) model by loosening the assumption of
constant pricing rule. By taking  the effect insider's strategy has
on pricing rule into consideration, we present three models vary
with different risk attitudes of the insider.

Model 1 shows the risk-averse insider case, where the insider
primarily maximizes the guaranteed profits. In equilibrium, the
insider trades in a way similar to that produced by the ``rat race"
effect once depicted by Holden and Subrahamanyam (1992) and Foster
and Viswanathan (1996). When trades happens more and more
frequently, the insider exploits the private information at an
exponential speed, yielding a strong-form efficient market within
any positive time. Model 2 focuses on the risk-neutral insider case,
where the insider primarily maximizes the ex ante expectation of
profits. In the limit as period number goes to infinity, Model 2
obtains exactly the continuous equilibrium in Kyle (1985). Model 3
presents the risk-seeking insider case, where the insider primarily
maximizes the risky profits. The risk-seeking behavior motivates
results similar to those by ``waiting game" effect in Foster and
Viswanathan (1996) with large trades  postponed and most information
 revealed in the latter periods of trading.

Comparison among the three models shows: 1, As to the information
revealing speed, Model 1 is highest and Model 3 is lowest. This
means that the more risk-averse one is,  the faster she will exploit
her information advantage. 2, The coefficient of  trading intensity
that increasing through trades for all the three models, implies the
fact that the downward trend of effect current trading has on the
future with time going on is robust respect to various risk
attitudes. 3, The adverse selection is decreasing through trades in
Model 1 and 2 with a lower decreasing speed in Model 2, whereas it
is increasing in model 3. This means that the more one is
risk-seeking, the more likely relatively high adverse selection will
be aroused when trading ends up.

\appendix

\renewcommand\thesection{\appendixname~\Alph{section}}

\renewcommand\theequation{\Alph{section}.\arabic{equation}}
\section{}
\setcounter{equation}{0} \text{{\Large P}ROOF OF {\Large
P}ROPOSITION 1:}
 (\ref{eq.2.3}) gives \begin{align*} \lambda_{1}^{'}
 (\beta_{1})
=\frac{\Sigma_{0}[-\beta^2_{1}\Sigma_{0}+\sigma^{2}_{u}\Delta{t_{N}}]}{[\beta^{2}_{1}\Sigma_{0}+\sigma^2_{u}\Delta{t_{N}}]^2}.
\end{align*}
By the Proposition 1 in Huddart etal., (2001), we have $
\beta^2_{1}\Sigma_{0}<\sigma^2_{u}\Delta{t_{N}}. $
Hence, (\ref{eq.2.5}) holds. $ \hfill Q.E.D.$\\

\text{{\Large P}ROOF OF {\Large P}ROPOSITION  2}: Denote
$\widetilde{y}_{1}, \widetilde{ y}_{2} \cdots, \widetilde{y}_{N}$ as
the orthonormalization of sequence $y_{1}, y_{2} \cdots, y_{N}$.
Clearly, for each $n$, $\widetilde{y}_{n}=\beta_{n} (v-p_{n-1})
+u_{n}$. Thus by definitions, any of $\Sigma_{k}, \lambda_{k}, $ or
$p_{k}$ $ (k=1, \cdots, N.) $ cannot be affected by $b_{n} (y_{1},
\cdots, y_{n-1}) $ and $c_{n}$, since they are all determined by
$\widetilde{y}_{k}, k=1, \cdots, N$.

Proof of the other conclusions need backward induction. In the last
period, insider earns
\begin{equation*}
\begin{split}
&E[\pi_{N}|p_{1}, \cdots, p_{N-1}, v]
=E[x_{N} (v-p_{N}) |y_{1}, \cdots, y_{N-1}, v]\\
&=E[x_{N} (v-p_{N-1}-\lambda_{N}\widetilde{y_{N}}) |y_{1}, \cdots, y_{N-1}, v]\\
&=\beta_{N} (1-\lambda_{N}\beta_{N}) (v-p_{N-1}) ^2+ (b_{N}
 (y_{1}, \cdots, y_{N-1}) +c_{N}) (1-\lambda_{N}\beta_{N})
 (v-p_{N-1}).
\end{split}
\end{equation*}
Thus $E[\pi_{N}|p_{1}, \cdots, p_{N-1}, v]$ has a formulation as
(\ref{eq.2.7}) with
\begin{align*}
\alpha_{N-1}=\beta_{N} (1-\lambda_{N}\beta_{N}),\quad h_{N-1}=
(b_{N} (y_{1}, \cdots, y_{N-1}) +c_{N}) (1-\lambda_{N}\beta_{N}),
\quad \delta_{N-1}=0.
\end{align*}
If the  conclusions hold for  $n+1th$ period, that is, there exists
non-random real numbers $\alpha_{n}, h_{n}, \delta_{n}$, such that
\begin{align*} E (\sum^{N}_{k=n+1}\pi_{k}|p_{1}, p_{2}, \cdots,
p_{n}, v) =\alpha_{n} (v-p_{n}) ^2+h_{n} (v-p_{n}) +\delta_{n},
\end{align*}
and $b_{n+1} (y_{1}, y_{2}, \cdots, y_{n}), c_{n+1}$ can affect term
$h_{n}$, but cannot affect  terms $\alpha_{n}$ or $\delta_{n}$. Then
stepping back by one period, we have:
\begin{equation*}
\begin{aligned}
&E[\sum^{N}_{k=n}\pi_{k}|p_{1}, p_{2}, \cdots, p_{n-1}, v]
=E[\sum^{N}_{k=n+1}\pi_{k}+ (v-p_{n}) x_{n}|y_{1}, y_{2}, \cdots, y_{n-1}, v]\\
&=E[\alpha_{n} (v-p_{n-1}-\lambda_{n}\widetilde{y_{n}}) ^2+h_{n} (v-p_{n-1}-\lambda_{n}\widetilde{y_{n}}) +\delta_{n}+ (v-p_{n-1}-\lambda_{n}\widetilde{y_{n}}) x_{n}|y_{1}, y_{2}, \cdots, y_{n-1}, v]\\
&=[\alpha_{n} (1-\lambda_{n}\beta_{n}) ^2+\beta_{n}
 (1-\lambda_{n}\beta_{n}) ] (v-p_{n-1}) ^2+ (b_{n} (y_{1}, \cdots,
y_{n-1}) +c_{n}+h_{n}) (1-\lambda_{n}\beta_{n}) \\
 &\quad
 (v-p_{n-1}) +\delta_{n}+\alpha_{n}\lambda^{2}_{n}\sigma^2_{u}\Delta{t_{N}}.
\end{aligned}
\end{equation*}
Hence (\ref{eq.2.7}), with (\ref{eq.2.8}) - (\ref{eq.2.10}) holds
for $n$.  Thus, $b_{n} (y_{1}, \cdots, y_{n-1}) $ and $c_{n}$ cannot
affect $\alpha_{n-1} (v-p_{n-1}) ^2+\delta_{n-1}$ since they cannot
affect price $p_{n-1}$ or parameters $\alpha_{n}, \lambda_{n}$,
$\delta_{n}$ in the formulations (\ref{eq.2.8}) and
 (\ref{eq.2.10}).
 In conclusion, Proposition \ref{pro.2} holds.

$ \hfill Q.E.D.$\\

\text{{\Large P}ROOF OF {\Large P}ROPOSITION  3:} Firstly, from
(\ref{eq.3.10}), we have
 (\ref{eq.3.25}). Then substituting for $a_{n}$ and $a_{n-1}$ from(\ref{eq.3.25}) into (\ref{eq.3.8}),
and rearranging yields
\begin{align}\label{eq.A.10} b_{n-1}=2
 (\frac{1}{c^2_{n}+1}) ^{3/2} (\frac{c^2_{n-1}+1}{2-c^2_{n-1}})
 b_{n}.
\end{align}
Noting that both (\ref{eq.A.10}) and (\ref{eq.3.9}) are expressions
for $b_{n-1}$, their equivalency gives (\ref{eq.3.24}). Substituting
for $b_{n}$ and $b_{n-1}$ from (\ref{eq.3.24}) into (\ref{eq.3.9}),
and rearranging shows (\ref{eq.3.22}).

The  monotonicity is verified as follows. Firstly, the equation
 (\ref{eq.3.22}) may be written as
\begin{align}\label{eq.A.11}
 (1-c^2_{n-2}/2) (1+c^2_{n-1}) ^{1/2}\frac{c^2_{n-1}}{c^2_{n-2}}
=\frac{ (1+c^2_{n-1}) /c_{n-1}}{ (1+c^2_{n}) /c_{n}}.
\end{align}
A direct calculation shows $c_{N-1}>1>c_{N-2}>c_{N-3}$. Secondly,
suppose for a general $n$, the inequality
\begin{align}\label{eq.A.12}
c_{n-1}<c_{n}<1\end{align} holds.
 Then claim that $c_{n-2}<c_{n-1}$, otherwise $c_{n-2}\geq c_{n-1}$, a contradiction follows.
In fact, for  (\ref{eq.A.11}), by (\ref{eq.A.12}), $RHS>1$. But
\begin{align*}
LHS\leq
 (1-c^2_{n-1}/2) (1+c^2_{n-1})
 ^{1/2}\frac{c^2_{n-1}}{c^2_{n-2}}\leq1.
\end{align*}
Thus, equation (\ref{eq.A.11}) cannot hold.  In conclusion, by the
backward induction, the monotonicity (\ref{eq.3.23}) holds.

 $ \hfill Q.E.D.$\\

\text{{\Large P}ROOF OF {\Large T}HEOREM {\ref{theorem.2}}:} There
are three steps in this proof. In step one we show that for any
continuous time $t\in (0,1) $, in the limit as $N\rightarrow
\infty$, $c_{[Nt]}$ converges to zero and both $b_{[Nt]}$ and
$a_{[Nt]}$ converge to infinity. In step two, we give the
convergence speeds of $a_{[Nt]}, b_{[Nt]}, c_{[Nt]}$. In step three,
based on the former two steps, we give the characteristics of limit
behaviors for sequences $\{\Sigma_{n}\}, \{\lambda_{n}\},$ and
$\{\beta_{n}\}$ as
$N\rightarrow\infty$.\\

\emph{Step one}: This step employs a method once used in the limit
equilibrium in Holden and Subrahmanyam (1992). To emphasize the
dependence, denote $c_{n}$ in economy with $N$ periods by
$c^{N}_{n}$. As seen from (\ref{eq.3.22}) in Proposition 3, the
strategy coefficient $c^{N}_{n}$ depends only on the the remaining
opportunity number $N-n$, that is
\begin{align}\label{eq.A.09}
c^{N}_{n}=c^{N+k}_{n+k} \qquad  for\ any\ natural\ numbers\ N,k,
with\
 n\leq N.
\end{align}
Combining this fact with the monotonicity property (\ref{eq.3.23}),
we know
\begin{align}\label{eq.A.14}
c^{N}_{n}\ is\ a\ decreasing\ function\ of\ N-n,\ with\ n\leqslant
N-1.
\end{align}
Now, hold a particular time $t\in (0,1) $. For any subsequence of
$\{1,2,\cdots,n,\cdots\}$, we can choose a sub-subsequence from it,
denoted as $N_{1}, N_{2}, \cdots, N_{k}, \cdots$, such
that\footnote{ (\ref{eq.A.15}) can be satisfied since
$N_{k}-[N_{k}t]\geq (1-t) N_{k} \rightarrow \infty$ as
$k\rightarrow\infty$. }
\begin{align}\label{eq.A.15}
N_{k+1}-N_{k}>[N_{k+1}t]-[N_{k}t],
\end{align}
 then $c^{ N_{k} }_{[N_{k}t]}$ is decreasing with $k$ and hence
 has a limit as $k\rightarrow\infty$, denoted as $c_{t}$. For the same reason $c^{ N_{k} }_{[N_{k}t]-1}$
also has a limit, denoted as $c^{'}_{t}$. \emph{Claim that
$c_{t}=c^{'}_{t}$, and the proof follows.}  For holden $n_{0}$, as
$N\rightarrow\infty$, $c^{N}_{n_{0}}$ has a limit since it is
decreasing with $N$. That is, for any $\epsilon>0$, there exists
$N_{0}$, when $N,N^{'}>N_{0}$, we have
\begin{align}\label{eq.A.16}|c^N_{n_{0}}-c^{N'}_{n_{0}}|<\epsilon.
\end{align}
Accordingly, when $k$ is big enough, such that
$N_{k}-[N_{k}t]+n_{0}>N_{0},$ then (\ref{eq.A.09}) and
(\ref{eq.A.16}) give
\begin{align*}
&|c^{N_{k}}_{N_{k}t}-c^{N_{k}}_{[N_{k}t]-1}|
=|c^{N_{k}-[N_{k}t]+n_{0}}_{n_{0}}-c^{N_{k}-[N_{k}t]+1+n_{0}}_{n_{0}}|<\epsilon.
\end{align*}
From the arbitrariness of $\epsilon$, we have
$c_{t}=\lim\limits_{k\rightarrow\infty}c^{ N_{k} }_{[N_{k}t]}
=\lim\limits_{k\rightarrow\infty}c^{ N_{k} }_{[N_{k}t]-1}
=c^{'}_{t}.$ Hence, the claim has been proved. An examination of
this proof shows
$\lim\limits_{k\rightarrow\infty}c^{N_{k}}_{[N_{k}t]-2}=\lim\limits_{k\rightarrow\infty}c^{
N_{k}}_{[N_{k}t]-1} =\lim\limits_{k\rightarrow\infty}c^{
N_{k}}_{[N_{k}t]} =c_{t}.$
 Thus, when $N\rightarrow\infty$,
substituting $c_{n}=c_{n-1}=c_{n-2}=c_{t}$ into equation
(\ref{eq.3.22}), we have\begin{align*}
 (1+c^2_{t}) (2-c^2_{t}) c^3_{t}=2 (1+c^2_{t}) ^{1/2} c^3_{t} \end{align*}
with a unique real root $c_{t}=0$. In conclusion,
\begin{align}\label{eq.A.19}\lim\limits_{k\rightarrow\infty}c^{ N_{k}
}_{[N_{k}t]}=0.\end{align} By the arbitrariness of subsequence we
choose, (\ref{eq.A.19}) implies$ \lim\limits_{N\rightarrow\infty}c^{
N }_{[Nt]}=0,$ i.e., (\ref{eq.3.26}) holds. Further, from
(\ref{eq.3.26}), (\ref{eq.3.27}) and (\ref{eq.3.28})
can be easily obtained from (\ref{eq.3.24}) and (\ref{eq.3.25}).\\

\emph{Step two}: In this step, it is crucial to find the speed of
$c^{ (N) }_{[Nt]}$ converging  to zero. To do this, take the
function \emph{ln} on both sides of (\ref{eq.3.22}), and rearrange,
yielding
\begin{align}\label{eq.A.21}
3\ln c_{n-1}-2\ln c_{n-2}-\ln c_{n}=\frac{1}{2}\ln (1+c^2_{n-1})
-\ln (1+c^2_{n}) -\ln (1-\frac{1}{2}c^2_{n-2}).
\end{align}
Define \begin{align}\label{eq.A.180}d_{n}=c^2_{n},\
q_{n}=\ln{c_{n}}.\end{align} Then by Taylor expansion near zero $\ln
(1+x) =x-\frac{1}{2}x^2+O (x^3) $, (\ref{eq.A.21}) can be written as
\begin{equation}
\begin{aligned}\label{eq.A.22}
3q_{n-1}-2q_{n-2}-q_{n}&=\frac{1}{2} (d_{n-1}-\frac{1}{2}d^2_{n-1})
- (d_{n}-\frac{1}{2}d^2_{n})
- (-\frac{1}{2}d_{n-2}-\frac{1}{8}d^2_{n-2}) \\
&\quad+O (d^3_{n-2}+d^3_{n-1}+d^3_{n}).
\end{aligned}
\end{equation}
Let
\begin{equation}\label{eq.A.23}
\begin{aligned}
&d_{n}=d_{t}\Delta{t}_{N}^{\beta},\\
 &d_{n-1}= (d_{t-\Delta{t}_{N}}\Delta{t}_{N}^{\beta})
= (d_{t}-d^{'}_{t}{\Delta{t}_{N}}) \Delta{t}_{N}^{\beta}+O (\Delta{t}_{N}^{2+\beta}), \\
 &d_{n-2}=d_{t-2\Delta{t}_{N}}\Delta{t}_{N}^{\beta}) =
 (d_{t}-2d^{'}_{t}{\Delta{t}_{N}}) \Delta{t}_{N}^{\beta}+O
 (\Delta{t}_{N}^{2+\beta}),
\end{aligned}
\end{equation}
in which $\beta$ is a parameter to be fixed later. By Taylor
expansion near zero $\ln (1+x) =x+O (x^2) $, (\ref{eq.A.180}) and
(\ref{eq.A.23}) give
\begin{equation}\label{eq.A.24}
\begin{aligned}
&q_{n}=\frac{1}{2}\ln d_{n}=\frac{1}{2}\beta\ln
\Delta{t}_{N}+\frac{1}{2}\ln d_{t}, \\
&q_{n-1}=\frac{1}{2}\ln d_{n-1} =\frac{1}{2}\beta\ln
\Delta{t}_{N}+\frac{1}{2}\ln d_{t}-\frac{1}{2}\frac{d'_{t}}{d_{t}}\Delta{t}_{N}+O (\Delta{t}_{N}^2), \\
&q_{n-2}=\frac{1}{2}\ln{d_{n-2}} =\frac{1}{2}\beta\ln
\Delta{t}_{N}+\frac{1}{2}\ln
d_{t}-\frac{d'_{t}}{d_{t}}\Delta{t}_{N}+O (\Delta{t}_{N}^2).
\end{aligned}
\end{equation}
Substituting (\ref{eq.A.24}) into LHS of (\ref{eq.A.22}) and
rearranging yields
\begin{align*}
LHS=\frac{1}{2}\frac{d'_{t}}{d_{t}}\Delta{t}_{N}+O
(\Delta{t}_{N}^2).
\end{align*}
Similarly,
\begin{equation*}
\begin{aligned}
RHS=-\frac{3}{2}d'_{t}\Delta{t}_{N}^{1+\beta}+\frac{3}{8}d^2_{t}{\Delta{t}_{N}}^{2\beta}
+O (\Delta{t}_{N}^{2+\beta})+O
 (\Delta{t}_{N}^{3\beta}).
\end{aligned}
\end{equation*}
Equating LHS to RHS yields $ \beta=\frac{1}{2}, \
 \frac{1}{2}\frac{d'_{t}}{d_{t}}=\frac{3}{8}d^2_{t}.$
Combining this differential equation with the terminal value
$d_{t}|_{t=1}=\infty,$ we get $d_{t}= (\frac{2}{3}\frac{1}{1-t})
^{1/2}.$ In conclusion, $\frac{d_{[Nt]}}{
\Delta{t}_{N}^{1/2}}\rightarrow{d_{t}}$ $(N\rightarrow\infty)$, and
equivalently,
\begin{align*}\frac{c_{[Nt]}}{ \Delta{t}_{N}^{1/4}}=\frac{d^{1/2}_{[Nt]}}{ \Delta{t}_{N}^{1/4}}\rightarrow (\frac{2}{3}\frac{1}{1-t}) ^{1/4} \quad
 (N\rightarrow\infty). \end{align*}
 This is exactly (\ref{eq.3.29}).
 (\ref{eq.3.30}) and (\ref{eq.3.31}) can be achieved from combining
 (\ref{eq.3.29}) (\ref{eq.3.24}) and (\ref{eq.3.25}).
For holden n, the limits of $c_{n}, b_{n}, a_{n}$ as
$N\rightarrow\infty$ correspond to $t=0$ in
(\ref{eq.3.29})-(\ref{eq.3.31}) respectively since
$n/N\rightarrow0$.\\

\emph{Step three}: Now, we take the limits of sequences
$\{\Sigma_{n}\}, \{\lambda_{n}\}$ and $\{\beta_{n}\}$ as
$N\rightarrow\infty$ using the former preliminary results.  From
(\ref{eq.3.17}) and (\ref{eq.3.7}),
\begin{equation*}\begin{aligned} \Sigma_{n}
=\frac{1}{ (1+c^2_{n}) (1+c^2_{n-1}) \cdots (1+c^2_{2})
 (1+c^2_{1}) }\Sigma_{0}=exp\{-\sum^n\limits_{i=1}\ln (1+c^2_{i})
\}\Sigma_{0}.
\end{aligned}\end{equation*}
Thus, for $t\in (0, 1) $, the monotonicity property of sequence
$\{c_{n}\}$ implies (\ref{eq.3.33}). Note that $c^{2}_{1}$ has order
$O (\Delta{t}_{N}^{1/2}) $ from which we know as
$\Delta{t}_{N}\rightarrow 0$, $exp\{-ln (1+c^2_{1})
\Delta{t}^{-1}_{N}\} \rightarrow 0$ and thus (\ref{eq.3.33}) gives
(\ref{eq.3.32}).
 From (\ref{eq.3.16}) and
 (\ref{eq.3.7}),
$$\lambda_{n}=\frac{c_{n}\Sigma^{1/2}_{n-1}}{ (1+c^2_{n}) \sigma_{u}
\Delta t_{N}^{1/2}},$$ implying (\ref{eq.3.34}) and (\ref{eq.3.35}).
Results about sequence $\{\beta_{n}\}$ can be deduced easily from
(\ref{eq.3.7}).
$ \hfill Q.E.D.$\\

\text{PROOF OF THEOREM {\ref{theorem.3}}}:  In the last period,
solutions are the same as those in Model 1, since for any strategy
taken from $X_{N}$, insider's guaranteed profits are zero, and thus
the three models degenerate to the same one.

Suppose in the $n+1th$ period, conclusions hold. Then, if the
informed submission in nth period is $\beta_{n} (v-p_{n-1}) $, the
profits accumulated from nth to the last period would be
\begin{align*}
&E[\sum^N_{k=n}\pi_{k}|p_{1}, \cdots, p_{n-1}, v]=\alpha_{n-1}
 (v-p_{n-1}) ^2+\delta_{n-1}
\end{align*} with
\begin{equation*}
\begin{aligned}
\alpha_{n-1}=\alpha_{n} (1-\lambda_{n}\beta_{n}) ^2+\beta_{n}
 (1-\lambda_{n}\beta_{n}), \quad
\delta_{n-1}=\delta_{n}+\alpha_{n}\lambda^2_{n}\sigma^2_{u}\Delta{t_{N}}.
\end{aligned}
\end{equation*}
Note that relationships (\ref{eq.3.19}) and (\ref{eq.3.20}) in Model
1 still hold in Model 2. Thus, by the law of iterated expectation,
\begin{align*}
E (\sum^N_{k=n}\pi_{k}) =
\frac{b_{n}\Sigma^{1/2}_{n-1}\sigma^2_{u}\Delta{t_{N}}}{ (
\beta^2_{n}\Sigma_{n-1}+\sigma^2_{u}\Delta{t_{N}}) ^{1/2}}
+\frac{\beta_{n}\Sigma_{n-1}\sigma^2_{u}\Delta{t_{N}}}{\beta^2_{n}\Sigma_{n-1}+\sigma^2_{u}\Delta{t_{N}}}
+a_{n}\sigma^2_{u}\Delta{t_{N}}
 (\frac{\Sigma_{n-1}}{\beta^2_{n}\Sigma_{n-1}+\sigma^2_{u}\Delta{t_{N}}})
 ^{1/2}.
\end{align*}
A tedious calculation gives
\begin{align*}
\frac{dE (\sum^N_{k=n}\pi_{k}) }{d\beta_{n}}=\frac{- (a_{n}+b_{n})
\beta_{n}\Sigma^{3/2}_{n-1}\sigma^2_{u}\Delta{t}_{N}}{
(\beta^2_{n}\Sigma_{n-1}+\sigma^2_{u}\Delta{t}_{N}) ^{3/2}}
+\frac{\Sigma_{n-1}\sigma^4_{u}{\Delta{t}^2_{N}}-\beta^2_{n}\Sigma^2_{n-1}\sigma^2_{u}\Delta{t}_{N}}
{ (\beta^2_{n}\Sigma_{n-1}+\sigma^2_{u}\Delta{t}_{N}) ^2}.
\end{align*}
Accordingly, the FOC yields \begin{align}\label{eq.A.37}
\beta_{n}=c_{n}\sigma_{u} \Delta t_{N}^{1/2}
\Sigma_{n-1}^{-1/2}\end{align}
 with $c_{n}$ satisfying
(\ref{eq.4.10}). While the SOC requires:
\begin{equation*}
\begin{aligned}
&\frac{d^{2}E (\sum^N\limits_{k=n}\pi_{k})
}{d{\beta_{n}}^2}=-(a_{n}+b_{n}) \Sigma^{1/2}_{n-1}
\sigma^2_{u}\Delta{t}_{N}
 \frac{ (-2\beta^2_{n}\Sigma^2_{n-1}+\Sigma_{n-1}\sigma^2_{u}\Delta{t}_{N}) }
 { (\beta^2_{n}\Sigma_{n-1}+\sigma^2_{u}\Delta{t}_{N}) ^{5/2}}
+\frac{2\beta^3_{n}\Sigma^3_{n-1}\sigma^2_{u}\Delta{t}_{N}-6\beta_{n}\Sigma^2_{n-1}\sigma^4_{u}{\Delta{t}_{N}}^2}{
(\beta^2_{n}\Sigma_{n-1}+\sigma^2_{u}
\Delta{t}_{N}) ^3}\\
&\quad<0.
\end{aligned}
\end{equation*}
Substituting (\ref{eq.A.37}) into the above yields
\begin{align}\label{eq.A.40}
 - (a_{n}+b_{n}) (-2c^2_{n}+1)
 (c^2_{n}+1) ^{-\frac{5}{2}}  +  (2c^3_{n}-6c_{n}) (c^2_{n}+1) ^{-3} <0.\end{align}
  Combined with (\ref{eq.4.10}), (\ref{eq.A.40}) can be satisfies by $c_{n}>0$ in our equilibrium.
$ \hfill Q.E.D.$\\

\text{{\Large P}ROOF OF {\Large P}ROPOSITION \ref{pro.4}}: From
(\ref{eq.4.10}), we have,
\begin{align}\label{eq.A.41}
b_{n}=\frac{1-c^2_{n}}{c_{n} (1+c^2_{n}) ^{1/2}}-a_{n}.
\end{align}
Substituting (\ref{eq.A.41}) into (\ref{eq.4.8}) yields
\begin{align}\label{eq.A.410}
a_{n-1}=a_{n} (\frac{1}{1+c_{n}^2}) ^{3/2}+\frac{ (1-c_{n}^2)
c_{n}}{ (1+c_{n}^2) ^2}.
\end{align}
 The sum of equation (\ref{eq.4.9}) plus (\ref{eq.A.410}) gives
\begin{align}\label{eq.A.42}
a_{n-1}+b_{n-1}= (a_{n}+b_{n}) (\frac{1}{1+c^2_{n}})
^{3/2}+\frac{2c_{n}}{ (1+c^2_{n}) ^2}. \end{align} Substituting for
$a_{n}+b_{n}$ and $a_{n-1}+b_{n-1}$ from (\ref{eq.4.10}) into
(\ref{eq.A.42}) and rearranging yields (\ref{eq.4.11}). To show the
monotonicity, change
 (\ref{eq.4.11}) to
\begin{align}\label{eq.A.43}
\frac{1-c^2_{n-1}}{c_{n-1} (1+c^2_{n-1}) ^{1/2}}=\frac{1}{c_{n}
 (1+c^2_{n}) }.
\end{align}
Since the LHS of (\ref{eq.A.43}) is decreasing with $c_{n-1}$, and
the RHS is decreasing with $c_{n}$, $c_{n-1}$ can be considered as
an increasing function of $c_{n}$. Combined with another fact
$c_{N-1}<c_{N}$, this fact implies $c_{N}>c_{N-1}>\cdots,
>c_{2}>c_{1}.$ $\hfill Q.E.D.$\\

\text{{\Large P}ROOF OF {\Large T}HEOREM 4}: The proof of
 (\ref{eq.4.13}) - (\ref{eq.4.15}) is similar to that in the step one of
Theorem 2 and is omitted. Take \emph{ln} on both sides of
 (\ref{eq.4.11}), we have
\begin{align}\label{eq.A.44}
\ln (1-c^2_{n-1}) +\ln c_{n}+ \ln (1+c^2_{n}) =\ln
c_{n-1}+\frac{1}{2}\ln (1+c^2_{n-1}). \end{align} By Taylor
expansion near zero $\ln (1+x) =x-\frac{1}{2}x^2+O (x^3)$ and let
$c_{n}=c_{t}\Delta{t}^{\alpha},
c_{n-1}=c_{t-\Delta{t}}\Delta{t}^{\alpha}=(c_{t}-c^{'}_{t}\Delta{t}_{N}+O
(\Delta{t}_{N} ^2))\Delta{t}^{\alpha},$ (\ref{eq.A.44}) gives
\begin{equation*}
\begin{aligned}&-\frac{3}{2}[
 (c_{t}-c^{'}_{t}\Delta{t}_{N}+O (\Delta{t}_{N} ^2))
^2\Delta{t}_{N}^{2\alpha}]-\frac{1}{4} (c_{t}-c^{'}_{t}+O
 (\Delta{t}_{N}) ^2) ^4\Delta{t}_{N}^{4\alpha}
+c^2_{t}\Delta{t}_{N}^{2\alpha}-\frac{1}{2}c^4_{t}\Delta{t}_{N}^{4\alpha}+O (\Delta{t}_{N} ^{6\alpha}) \\
&=-\frac{c^{'}_{t}}{c_{t}}\Delta{t}_{N}+\frac{O (\Delta{t}_{N}^2)
}{c_{t}}-\frac{1}{2}
 (-\frac{c^{'}_{t}}{c_{t}}\Delta{t}_{N}+\frac{O (\Delta{t}_{N}^2) }{c_{t}}) ^2+O (-\frac{c^{'}_{t}}{c_{t}}\Delta{t}_{N}+\frac{O (\Delta{t}_{N}^2) }{c_{t}}) ^3+O
 (\Delta{t}_{N} ^{6\alpha})
\end{aligned}
\end{equation*}
which implies $\alpha=\frac{1}{2}$,\ $
 \frac{1}{2}c^2_{t}=\frac{c^{'}_{t}}{c_{t}}.$ Combined with the terminal condition
 $c_{t}|_{t=1}=\infty$, we have
$c_{t}=\frac{1}{ (1-t) ^{1/2}},$ i.e., (\ref{eq.4.16}) has been
proved.

Now to show the limit result for sequence $\{b_{n}\}$,  let $
b_{n}=b_{t}{\Delta{t}_{N}^\alpha},\
b_{n-1}=b_{t-\Delta{t}_{N}}{\Delta{t}_{N}^\alpha}=(b_{t}-b^{'}_{t}\Delta{t}_{N}+O
(\Delta{t}_{N}^2)){\Delta{t}_{N}^\alpha} $, and by Taylor expansions
near zero $(1+x) ^{k}=1+kx+O (x^2)$, (\ref{eq.4.9}) gives
\begin{align*}
 (-b^{'}_{t}+\frac{3}{2}b_{t}c^2_{t}) \Delta{t}_{N}^{1+\alpha}-c_{t}\Delta{t}_{N}^{1/2}+O (\Delta{t}_{N} ^{3/2})+O (\Delta{t}_{N} ^{2+\alpha})
 =0.
\end{align*}
Thus, $\alpha=-\frac{1}{2},$\
$-b^{'}_{t}+\frac{3}{2}c^2_{t}b_{t}-c_{t}=0.$ Since
$b_{t}|_{t=1}=\infty$, (\ref{eq.4.17}) holds. Combined with
 (\ref{eq.4.16}) and (\ref{eq.4.17}), (\ref{eq.4.10}) gives
 (\ref{eq.4.18}).

At last,  show the other parameters' limit results
 (\ref{eq.4.19}) - (\ref{eq.4.21}). As  in Model 1, we
have a difference equation for sequence $\{\Sigma_{n}\}:$ $
\Sigma_{n-1}-\Sigma_{n}=\Sigma_{n}c^2_{n},$ which implies that the
continuous version of sequence $\{\Sigma_{n}\}$ satisfies
$-d\Sigma_{t}=\Sigma_{t}\frac{1}{1-t}dt$ and thus (\ref{eq.4.19})
holds. (\ref{eq.4.20}) and
(\ref{eq.4.21}) can be obtained similarly to those in Model 1.$ \hfill Q.E.D.$\\

\text{{\Large P}ROOF OF {\Large T}HEOREM \ref{theorem.5}}:
 Suppose in equilibrium, in the $n+1$th period, the conclusions
 hold.
Then strategy $\beta_{n} (v-p_{n-1}) $ in the nth period implies
\begin{align*}
&E[\sum^N_{k=n}\pi_{k}|p_{1}, \cdots, p_{n-1}, v]=\alpha_{n-1}
 (v-p_{n-1}) ^2+\delta_{n-1}
\end{align*} with
\begin{equation*}
\begin{aligned}
\alpha_{n-1}=\alpha_{n} (1-\lambda_{n}\beta_{n}) ^2+\beta_{n}
(1-\lambda_{n}\beta_{n}), \
\delta_{n-1}=\delta_{n}+\alpha_{n}\lambda^2_{n}\sigma^2_{u}\Delta{t_{N}}.
\end{aligned}
\end{equation*}
The following relationship still holds:
\begin{align}\label{eq.A.53}
\alpha_{n-1} (\beta_{n})
=b_{n}\sigma_{u}{{\Delta{t}}^{1/2}_{N}}\Sigma_{n}^{-1/2}
 (\frac{\sigma^2_{u}\Delta{t}_{N}}{\beta^2_{n}\Sigma_{n-1}+\sigma^2_{u}\Delta{t}_{N}})
^{3/2}+
\frac{\beta_{n}\sigma^2_{u}\Delta{t}_{N}}{\beta^2_{n}\Sigma_{n-1}+\sigma^2_{u}\Delta{t}_{N}}.
\end{align}
Thus the FOC   yields
$\beta_{n}=c_{n}\sigma_{u}{\Delta{t}}^{1/2}_{N}\Sigma_{n-1}^{-1/2}$
with $c_{n}$ satisfying (\ref{eq.5.10}), and the SOC is
\begin{align}\label{eq.A.54} -3b_{n}- ({c_{n}}^2+1)
^{-1/2}c_{n} (1+3c^2_{n}) <0.\end{align} Combined   with
 (\ref{eq.5.10}), (\ref{eq.A.54}) is satisfied since $c_{n}>0$ in equilibrium. $ \hfill Q.E.D.$\\

\text{{\Large P}ROOF OF {\Large P}ROPOSITION \ref{pro.5}}: Firstly,
(\ref{eq.5.13}) yields from  (\ref{eq.5.10}). Then substituting for
$b_{n}$ and $b_{n-1}$ from (\ref{eq.5.13}) into
 (\ref{eq.5.9}) and rearranging  yields
(\ref{eq.5.11}). Denote the LHS of (\ref{eq.5.11}) as $f_{1}
(c_{n-1}) $ RHS of it as $f_{2} (c_{n}) $. Calculations show

\begin{align*}
\frac{df_{1} (c_{n-1}) }{dc_{n-1}}<0,\quad \frac{df_{2} (c_{n})
}{dc_{n}}<0.
\end{align*}
Thus, $c_{n-1}$,  as a function of $c_{n}$, is increasing with it.
Combined this fact with
$c_{N}>c_{N-1}$, (\ref{eq.5.12}) holds.$ \hfill Q.E.D.$\\

\text{{\Large P}ROOF OF {\Large T}HEOREM \ref{theorem.6}}: Firstly,
 (\ref{eq.5.11}) can be changed to \begin{align}\label{eq.A.56}
 (c^2_{n-1}+1) ^{1/2} (1-c^2_{n-1}) (c^2_{n}+1) c_{n}= (1+2c^2_{n})
 c_{n-1}.
\end{align}
Let $c_{n}=c_{t}{\Delta{t}}^{\alpha}, \quad
c_{n-1}=c_{t-\Delta{t}}{\Delta{t}}^{\alpha}=(c_{t}-c^{'}_{t}\Delta{t^{\alpha}_{N}}){\Delta{t^{\alpha}_{N}}}+O({\Delta{t^{2+\alpha}_{N}}}).$
Then (\ref{eq.A.56}) implies
\begin{align*}{\Delta
t}^{\alpha}[-\frac{3}{2}c^3_{t}{\Delta{t}}^{2\alpha}+c'_{t}\Delta{t}]+O
 (\Delta{t}^{5\alpha})+O(\Delta{t}^{3\alpha+1}) +O (\Delta{t}^{2+\alpha})
 =0,
\end{align*}
from which, we have $\alpha=\frac{1}{2}, \
 -\frac{3}{2}c^3_{t}+c'_{t}=0.$
Combined with the terminal value $c_{t}|_{t=1}=\infty$, we know
$c_{t}=\frac{\sqrt{3}}{3}\frac{1}{ (1-t) ^{1/2}},$ thus
(\ref{eq.5.17}) is proved.

Now,  show the limit results for sequences $\{b_{n}\},\{a_{n}\}$.
From
 (\ref{eq.5.13})  and (\ref{eq.5.17}), we have limit result for $\{b_{n}\}$, i.e.,
 (\ref{eq.5.18}). Further,
  (\ref{eq.5.8}) can be changed to  \begin{align*}\label{eq.A.59} a_{n-1}
 (c^2_{n}+1) ^{3/2}=a_{n} (c^2_{n}+1) +b_{n}c^2_{n}.
\end{align*}
Let $ a_{n}=a_{t}{\Delta{t}}^{a}\quad
a_{n-1}=a_{t-\Delta{t}}{\Delta{t}}^{a}=(a_{t}-a^{'}_{t}\Delta{t}){\Delta{t}}^{\alpha}+O({\Delta{t}}^{2+\alpha}).
$ Similarly,
\begin{align*}
-a'_{t}{\Delta{t}}^{1+a}+\frac{1}{2}a_{t}c^2_{t}{\Delta{t}}^{1+a}+O
 ({\Delta{t}}^{2+a}) =\frac{\sqrt{3}}{9}\frac{1}{ (1-t)
^{1/2}}{\Delta{t}}^{1/2}.
\end{align*}
Which implies $a=-\frac{1}{2},\
-a^{'}_{t}+\frac{1}{2}a_{t}c^2_{t}=\frac{\sqrt{3}}{9}\frac{1}{ (1-t)
^{1/2}},$ and thus, $a_{t}=\frac{\sqrt{3}}{6} (1-t) ^{1/2},$ i.e.,
 (\ref{eq.5.19}) is proved.
 Calculations for continuous versions of sequences
$\{\Sigma_{n}\}$, $\{\lambda_{n}\}$ and $\{\beta_{n}\}$ are similar
to those in the former two models.

\begin{center}
{REFERENCES}
\end{center}

\footnotesize{ \noindent\hangafter1\hangindent1.5em  {\normalsize
A}DMATI, {\normalsize A}., AND {\normalsize P}FLEIDERER (1988): ``A
Theory of Intraday Patterns: Volume and Price Variability,"
 \emph{Review of Financial Studies}, 1, 3--40.

\noindent\hangafter1\hangindent1.5em   {\normalsize B}ARUCH,
{\normalsize S}. (2002): ``Insider Trading and Risk Aversion,"
\emph{Journal of Financial Markets}, 5, 451--464.

\noindent\hangafter1\hangindent1.5em  {\normalsize C}ALDENTEY
{\normalsize R}., AND {\normalsize E}. {\normalsize S}TACCHETTI
 (2010): ``
 Insider Trading With A Random Deadline," \emph{Econometrica}, 78, 245--283.

\noindent\hangafter1\hangindent1.5em  {\normalsize C}HAU,
{\normalsize M}., AND {\normalsize D}. {\normalsize V}AYANOS (2008)
: ``Strong-Form Efficiency with Monopolistic Insiders," \emph{The
Review of Financial Studies,} 21, 2275--2306.

\noindent\hangafter1\hangindent1.5em {\normalsize F}AMA,
{\normalsize E}. (1970): ``Efficient Capital Markets: A Review of
Theory and Empirical Work," \emph{The Journal of Finance }, 25,
383--471.

\noindent\hangafter1\hangindent1.5em  {\normalsize F}OSTER,
{\normalsize F}. AND {\normalsize S}. {\normalsize V}ISWANATHAN
 (1996): ``Strategic Trading When Agents Forecast the Forecast of
Others," \emph{The Journal of Finance}, 51, 1437--1478.

\noindent\hangafter1\hangindent1.5em {\normalsize H}UDDART,
{\normalsize S}., {\normalsize J}.{\normalsize H}UGHES AND
{\normalsize C}. {\normalsize L}EVINE (2001): ``Public Disclosure
and Dissimulation of Insider Traders," \emph{ Econometrica}, 69,
665--681.

\noindent\hangafter1\hangindent1.5em   {\normalsize H}OLDEN,
{\normalsize C}., AND {\normalsize A}. {\normalsize S}UBRAHMANYAM
 (1992): ``Long-Lived Private Information and Imperfect Competition,"
\emph{Journal of Finance }, 47, 247--270.

\noindent\hangafter1\hangindent1.5em
---\hskip-1pt---\hskip-1pt---\   (1994): ``Risk Aversion, Imperfect
Competition and Long Lived Information," \emph{Economics Letters},
44, 181--190.

\noindent\hangafter1\hangindent1.5em  {\normalsize H}UDDART,
{\normalsize S}., {\normalsize J}. {\normalsize H}UGHES AND
{\normalsize M}. {\normalsize W}ILLIAMS (2010): ``Pre-Announcement
of Insiders' Trades," Working paper, Pennsylvania State University.

\noindent\hangafter1\hangindent1.5em  {\normalsize K}ALLENBERG,
{\normalsize O}. (2002): \emph{Foundations of Modern Probability}.
2nd ed. New York: Springer-Verlag.

\noindent\hangafter1\hangindent1.5em  {\normalsize K}YLE,
{\normalsize A}. (1985): ``Continuous Auctions and Insider Trading,"
\emph{ Econometrica}, 53, 1315--1335.

\noindent\hangafter1\hangindent1.5em {\normalsize Z}HANG,
{\normalsize W}. (2004): ``Risk Aversion, Public Disclosure, and
Long-Lived Information," \emph{Economics Letters } 85, 327--334.}\\

\normalsize{\emph{Institute of Applied mathematics, Academy of
Mathematics and Systems Science, Chinese Academy of Sciences,
Beijing 100080, PR China.{\begin{center}and\end{center}}Department
of Business Statistics and Econometrics, Guanghua School of
Management, Peking University, Beijing, 100871, PR China.;
zhoudeqing@amss.ac.cn.}}

\end{document}